\documentclass[useAMS,usenatbib]{mn2e}

\usepackage{graphicx}
\usepackage{times}
\usepackage{natbib}
\usepackage{color}

\newcommand{\apjs}{ApJS}

\newcommand{\apjl}{ApJL}

\newcommand{\aap}{A \& A}

\newcommand{\enzo}{\it{\small ENZO}}

\usepackage[colorinlistoftodos]{todonotes}
 \begin{document}
 
\title[How complex is the Cosmic Web?] {How complex is the Cosmic Web?}
\author[F. Vazza]{F. Vazza$^{1,2,3}$\thanks{E-mail: franco.vazza2@unibo.it}\\
$^{1}$ Dipartimento di Fisica e Astronomia, Universit\'{a} di Bologna, Via Gobetti 92/3, 40121, Bologna, Italy\\
$^{2}$ Hamburger Sternwarte, Gojenbergsweg 112, 21029 Hamburg, Germany\\
$^{3}$Istituto di Radio Astronomia, INAF, Via Gobetti 101, 40121 Bologna, Italy}

\date{Received / Accepted}
\maketitle
\begin{abstract}
The growth of large-scale cosmic structure  is a beautiful exemplification of how complexity can emerge in our Universe, starting from simple initial conditions and simple physical laws. Using {\enzo} cosmological numerical simulations, I applied  tools from Information Theory (namely, "statistical complexity") to quantify the amount of complexity in the simulated cosmic volume, as a function of cosmic epoch and environment. This analysis can quantify  how much difficult to predict, at least in a statistical sense, is the evolution of the thermal, kinetic and magnetic energy of the dominant component of ordinary matter in the Universe (the intragalactic medium plasma). 
The most complex environment in the simulated cosmic web is generally found to be the periphery of large-scale structures (e.g. galaxy clusters and filaments), where the complexity is on average $\sim 10-10^2$ times larger than in more rarefied regions, even if the latter dominate the volume-integrated complexity of the simulated Universe.  
If the energy evolution of gas in the cosmic web is measured on a $\approx 100 $ $\rm kpc/h$ resolution and over a $\approx 200$ $\rm Myr$ timescale, its total complexity is the range of $\sim 10^{16}-10^{17} \rm ~bits$, with little dependence on the assumed gas physics, cosmology or cosmic variance. 

\end{abstract}

\label{firstpage} 

\section{Introduction}
\label{sec:intro}

The description of physical processes in Nature routinely faces the concept of "complexity", usually meant as the reason why achieving a self-consistent  description of many natural phenomena is often so challenging.  

However, Information Theory \citep[e.g.][for an inspiring introduction]{prokopenko2009information}, suggests that 
 not all systems whose evolution is complicated to compute or to predict shall be considered  truly {\it complex} in a physical sense. A purely random process, for example, is almost impossible to predict in detail, but in a statistical sense its evolution is 
trivial to compute. Conversely, a truly complex phenomenon requires a significant amount of information to predict its evolution in configuration space even in a statistical sense.  
 A complex dynamics is often found to emerge from a very limited set of (seemingly simple) initial conditions and physical laws \citep[e.g.][for a recent review]{Glattfelder2019}. 
 
 Astrophysics is no exception here.
 The Universe that astrophysicists routinely analyse gives a spectacular example of such emergence from simple initial conditions: somehow the Universe could  self-organize  on an enormous range of scales without any external intervention, transitioning from the  smoothest and simplest possible initial condition (a nearly scale-invariant background of matter fluctuations, $\delta \rho/\rho \leq 10^{-5}$, in an expanding spacetime, where $\rho$ is the matter density) to a majestic hierarchy of clustered sources, with $\delta \rho /\rho \geq 10^4-10^5$ density contrasts across scales of tens of $\sim ~\rm Mpc$ \citep[e.g.][]{PE99.2,2015eaci.book.....S}.  
Its observed  clustering properties cannot be easily understood or predicted from its main build blocks alone (say galaxies or dark matter halos), but have instead somehow emerged from the complex interplay between many components and many different scales of interaction. 

Therefore, the Universe perfectly fits into the standard definition of what a complex system is{\footnote{See https://complexityexplained.github.io for a recent public repository of resources and visualization tools to explore complexity in physics.}}, further motivating a first numerical analysis of where and how such complexity may have arisen. Applications of Information Theory to characterize the complexity of the observed distribution of galaxies and their bias with respect to the total matter distribution have been recently proposed \citep[][]{2016MNRAS.463.4239P}.\\

From the reductionist  perspective of computer simulations, the natural  emergence of "cosmic complexity"  is exemplified by the fact that a single random number of a few digits, combined with a source code that can be stored in a a few hundreds of Kilobytes of text (relying on a few more numerical libraries and compilers) can produce extremely rich  and vivid digital models of our Universe, which can only be stored by many tens of Terabytes of data (and counting).\\

The minimal numerical model for the formation of cosmic structures on a computer (which will also be the subject of this work) is made of  a set of nearly scale-invariant initial conditions for linear mass and velocity perturbations, an effective equation of state for the gas component, a budget for the relative energy density of dark matter, ordinary matter and dark energy, and numerical routines to solve for the effect of gravity, hydrodynamics in an expanding space.\\

Beside using this numerical model for testing various models for the evolution of interesting astrophysical objects  against observations (which invariably leads to major or minor revision of the assumed physical scenarios for the evolution of large scale structures and/or for the formation and growth of galaxies, in a continuous learning process)  the same  stream of simulated data can also be analyzed with a symbolic analysis, in to measure the internal symbolic "dialog" between the different algorithm components, that result into so rich  simulation outputs. \\

A first attempt to do this has been presented in \citet{va17info},  in which I applied methods derived from Information Theory to quantify for the first time the spatial distribution and time evolution of {\it statistical complexity} and {\it block entropy} in modern cosmological simulations\footnote{To the best of my knowledge, most of documented applications of Information Theory to astrophysics concern the reconstruction of sparse signals in real data \citep[e.g.]{ens09,ens11,ens13}, cosmology \citep[e.g.][]{2004PhRvL..92n1302H,2012PhRvD..86h3539L}, the analysis of clustering in extragalactic surveys \citep[e.g.][]{2013MNRAS.430.3376P,2015MNRAS.454.2647P}, or the evolution of compact stars  \citep[][]{deAvellar20121085}.}.
This approach makes it possible to quantify how many bits of information are necessary to predict the evolution of cosmic energy fields, considering each resolution element in the simulation as an independent "information processing device". In detail, in  \citet{va17info}  I focused on the high-resolution view 
 of one  massive galaxy cluster, measuring that the most complex behaviors are found in the peripheral cluster regions, where supersonic flows drive shocks and large energy fluctuations over a few tens of million years. At high numerical resolution, the amplification of magnetic fields by dynamo introduces more complexity, while non-gravitational physics involved in galaxy formation (e.g. radiative cooling and feedback from active galactic nuclei) adds significant complexity to the evolution of intracluster gas at most epochs.   I have also measured that the typical scale at which are "richest" of information, i.e. on which scales a bit of information gives the highest value in predicting capabilities,  is around
$\sim 10^2$ kpc, consistent with the fact these are typical scales of the largest turbulent and magnetic eddies in the simulated cosmic gas. \\

\begin{figure*}
\includegraphics[width=0.33\textwidth]{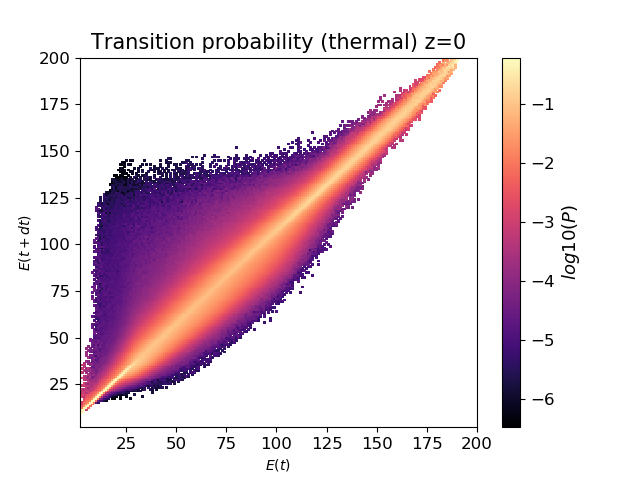}
\includegraphics[width=0.33\textwidth]{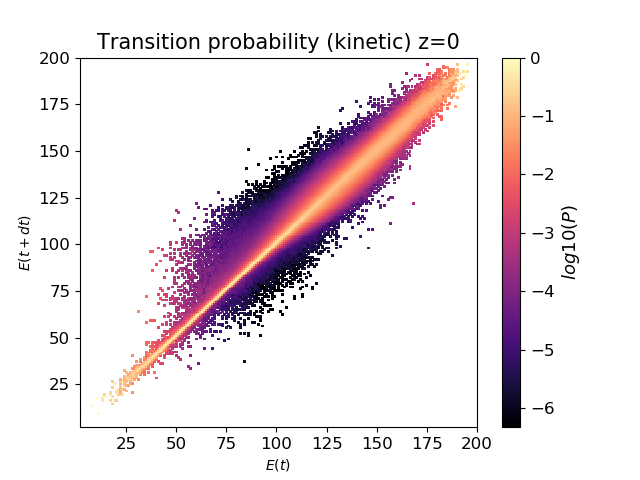}
\includegraphics[width=0.33\textwidth]{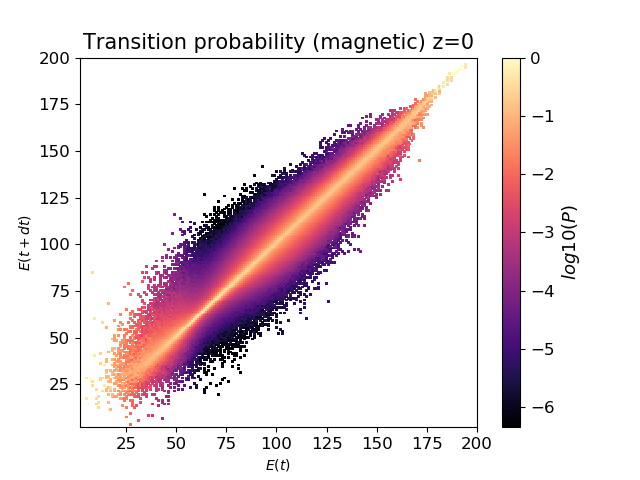}
\caption{Matrix of transition probabilities, $P_{xyz}$ between energy states at $t$ and $t+dt$, considering transitions of thermal (left), kinetic (centre) and magnetic (right) energy at $z=0.0$. The number of energy bins is $N_{\rm bin}=200$ in all cases, and the $dt=5$ timesteps here.}
\label{fig:prob}
\end{figure*}

\begin{table*}
\label{tab_cosmo}
\caption{List of cosmological {\enzo} simulations produced for this work.}
\centering \tabcolsep 5pt 
\begin{tabular}{c|c|c|c|c|c|c|c|c}
Run ID & Volume  & Resolution & $\Omega_M$ &  $\Omega_b$  &  $\Omega_\Lambda$ & $\sigma_8$ & $h$ & description\\
       &   [Mpc/h]    &   [kpc/h]     &       &      &     &     & [$100 ~\rm km/s/Mpc$] \\ \hline
Cosmo0 &  40 &  100 &  0.3 & 0.04 & 0.7 & 0.8 & 0.7&baseline \\
Cosmo0cool &  40 &  100 &  0.3 & 0.04 & 0.7 & 0.9 & 0.7  & cooling\\
Cosmo0AGN &  40 &  100 &  0.3 & 0.04 & 0.7 & 0.9 & 0.7  & cooling+AGN\\
Cosmo0AGN2 &  40 &  100 &  0.3 & 0.04 & 0.7 & 0.9 & 0.7  & cooling+AGN (high power)\\
Cosmo2 &  40 &  100 &  0.3 & 0.04 & 0.7 & 0.9 & 0.7  & high $\sigma$\\
Cosmo3 &  40 &  100 &  0.3 & 0.04 & 0.7 & 0.7 & 0.7  & low $\sigma$\\
Cosmo3A &  40 &  100 &  0.3 & 0.04 & 0.7 & 0.7 & 0.7  & cosmic variance\\
CosmoCDM &  40 &  100 &  0.96 & 0.04 & 0.0 & 0.43 & 0.7  & CDM\\
Cosmo4 &  40 &  100 &  0.308 & 0.0478 & 0.692 & 0.815 & 0.678 & Planck \\

\end{tabular}
\end{table*}

Building on the same methodology, in this new work I will discuss the analysis of complexity in a larger volume, simulated at a fixed spatial and mass resolution, and including several variations of the cosmological or physical parameters. This approach allows to address the simple question: how complex is the (simulated) cosmic web?

In detail, in Sec.\ref{sec:methods} I give the description of the set of cosmological simulations and on of the numerical algorithms to measure statistical complexity; in Sec.\ref{sec:res} I give the measured properties of complexity in the cosmic volume and as a function of cosmic time; in Sec.\ref{variations} I discuss the impact of algorithmic or physic variations with respect to the baseline model. Finally,  in the Section \ref{sec:conclusions} I summarize my results and apply them to estimate the total complexity of the cosmic web within the entire visible Universe, and I give some future perspectives on this new astrophysical line of research.

\begin{figure}
\includegraphics[width=0.48\textwidth]{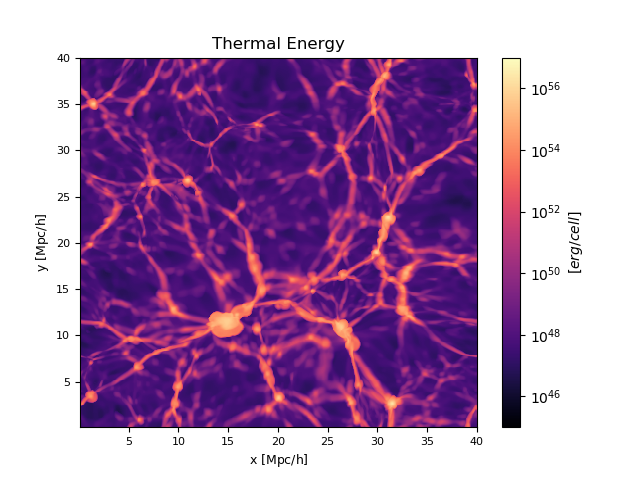}
\includegraphics[width=0.48\textwidth]{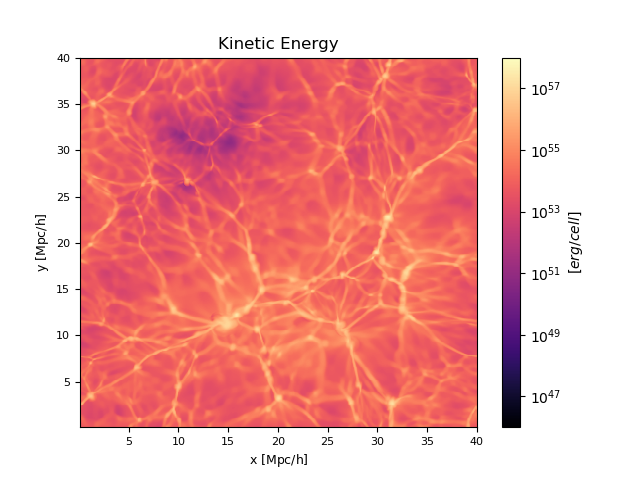}
\includegraphics[width=0.48\textwidth]{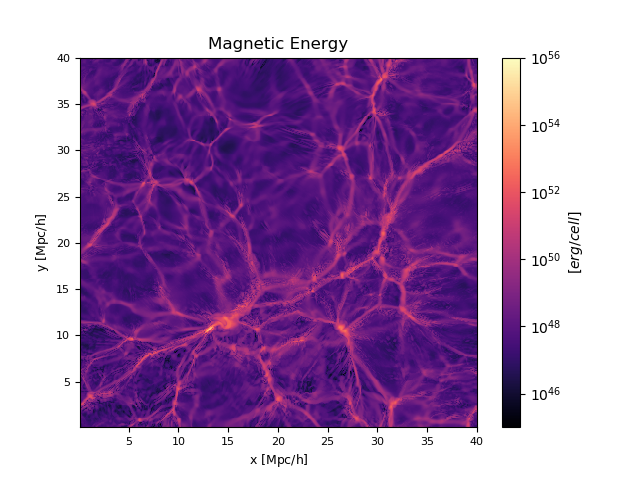}
\caption{Slice of thermal (top), kinetic (centre) and magnetic (bottom) energy at $z=0.0$, for the full $40 \times 40 \rm ~Mpc/h^2$ simulation and for a thickness of $100 \rm ~kpc/h$ along the line of sight. }
\label{fig:cut_fields}
\end{figure}

\begin{figure}
\includegraphics[width=0.48\textwidth]{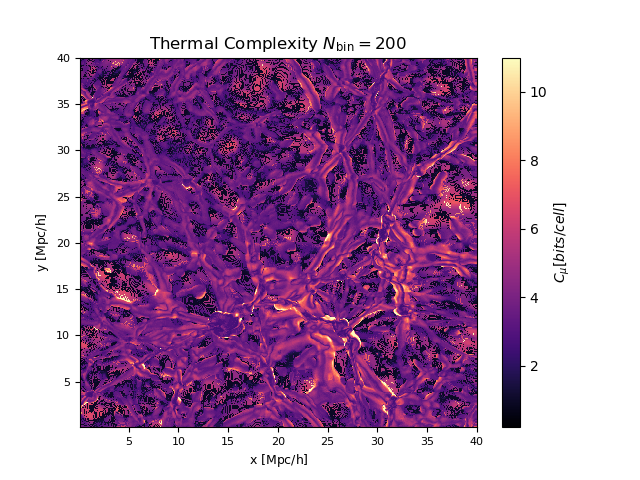}
\includegraphics[width=0.48\textwidth]{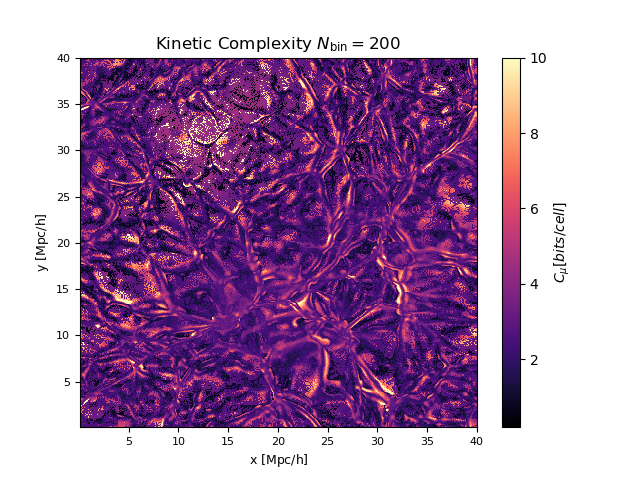}
\includegraphics[width=0.48\textwidth]{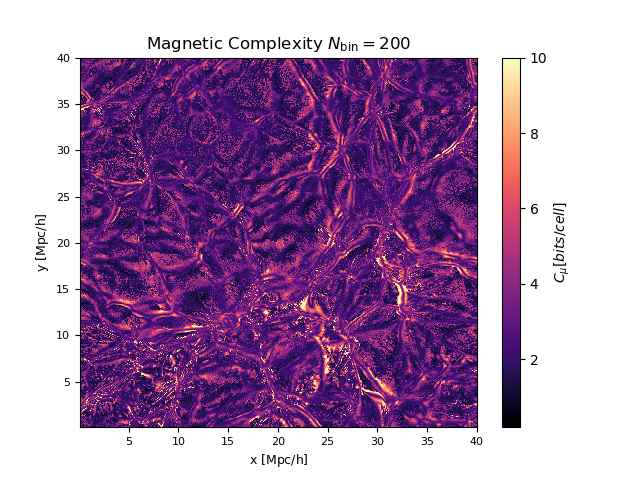}
\caption{Slice of thermal (top), kinetic (centre) and magnetic (bottom) statistical complexity at $z=0.0$, for the same selection of Fig.\ref{fig:cut_fields}.}
\label{fig:cut_info200}
\end{figure}

\section{Methods}
\label{sec:methods}
\subsection{Cosmological Simulations}
\label{subsec:sim}

I produced cosmological magneto-hydrodynamical (MHD)  simulations of  a cosmic volume of $40^3 \rm ~Mpc/h^3$ comoving, using a customized version of the publicly available {\enzo} code v.3.0 \citep[][]{enzo14} .   

One run (Cosmo0)
is devoted to simulate the volume with a fixed number of cells and dark matter particles ($400^3$) in a reference $\Lambda$CDM cosmology with parameters  $\Omega_{\rm b} = 0.04$, $\Omega_{\rm M} = 0.3$, $\Omega_{\Lambda}=0.7$,  a Hubble constant $H_0 = 70.0 \rm ~km/s/Mpc$ and a normalization for the primordial matter power spectrum of $\sigma_8=0.8$.  The starting redshift of the simulation was set to $z=30$ ($t_U\approx 0.1$ Gyr). 

With a set of additional runs I investigated variations on the above  "baseline" model (always keeping the $n^3=400^3$ configuration as fixed), in order to study the impact of cosmological parameters on complexity. In runs Cosmo2 and Cosmo3 I varied the power spectrum normalization ($\sigma_8=0.9$ or $\sigma_8=0.7$, respectively) while in Cosmo4 I adopted the  PLANCK cosmology \citep[][]{2016A&A...594A..13P}  with  $\Omega_{\rm b} = 0.0441$, $\Omega_{\rm M} = 0.258$,  $\Omega_{\Lambda} = 0.742$, a Hubble constant of $H_0 = 67.8$ km/sec/Mpc and  $\sigma_8=0.815$. 
For completeness, as an additional test I also considered a simpler Cold Dark Matter scenario (run CosmoCDM),  with parameters tuned to roughly reproduce the halo mass function of the baseline run, i.e. by assuming $\Omega_{\rm M}=1.0$, $\Omega_\Lambda=0.0$ and $\sigma_8=0.43$ (since the halo abundance scales with $\Omega_M \sigma_8^0.5$ \citealt[e.g.][]{2002ARA&A..40..539R}).
Finally, run Cosmo3A is the additional random realization of initial conditions using the same cosmology of run Cosmo3 (meant to investigate cosmic variance), while run Cosmo0cool,  Cosmo0AGN and Cosmo0AGN2 are resimulations of the same cosmology as in the baseline run, but including the effect of radiative gas cooling  and of the thermal and magnetic feedback by active galactic nuclei. In the latter case  I used a simplified sub-grid modelling approach optimized for the specific resolution used, with details given in  \citet{va13feedback},  in which a fixed thermal and magnetic power per event is imposed whenever the gas density is increased by radiative cooling exceeding a critical threshold ($\approx 10^{-2} \rm ~part/cm^3$). In this work I tested a fixed value of $10^{58} ~\rm erg$ (Cosmo0AGN) or $10^{59} ~\rm erg$ (Cosmo0AGN2) for each feedback event (with a fixed $10\%$ of it released in the form of magnetic energy, similar to \citealt{hack16}) which leads to a rather continuous or more impulsive feedback cycle in simulations, respectively.

A list of all runs analysed in this work is given in Tab.\ref{tab_cosmo}.  

To solve for the evolution of magnetic fields, all simulations employed the MHD scheme of Dedner \citep[][]{ded02} using the HLL solver {\enzo} \citet{wa09}.  The magnetic field in all runs is simply initialised by imposing a uniform $B_0=0.1 ~\rm nG$ (comoving) field strength along all magnetic field components at the start of the simulation. In radiative runs, however, the initial magnetic field is a factor $100$ lower, to accomodate for the additional release of  magnetic fields by AGN  as in  \citep{va17cqg}.\\

The choice of this constant spatial resolution (which is coarser than what modern simulations can typically afford) is motivated by the previous analysis of 
the "efficiency of prediction" in  \citet{va17info}, which concluded that this intermediate range of scales is the optimal one to study complexity in cosmology.
 In detail, the effiency of prediction  \citep[e.g.][]{2004PhRvL..93n9902S,prokopenko2009information} measures the scale at which making predictions about the future state of a physical system  gives the best "emergent" view, i.e. the optimal compromise between the perfect but trivial predictions that can be produced with a low-resolution version of data, and the purely random behaviour that can be typically measured on extremely small scales. In the case of the cosmic web, 
the first extreme is met on very small scales ($\ll 10~\rm kpc$) where plasma fluctuations get chaotic and, in real Universe, turbulence can be continuously fed by small scale instabilities and collision-less effects \citep[e.g.][]{sch05,bl11}, while the second extreme is met on very large scales ($\gg \rm ~Mpc$), on which the matter density contrast remains small and evolution can be computed with a linear approximation \citep{ze70}. This first study constrained the range of scales $\sim 50-200~ \rm kpc$ as the best to maximize the efficiency of prediction, as this is the range of scales at which  at which the most important physical interactions (e.g. accretion of subclumps and cluster satellites, injection of turbulence and shocks, growth of magnetic field eddies, etc) takes place. 

Therefore, while in \citet{va17info} I degraded the resolution of the simulation in order to measure complexity on the best scale, for this work I directly generated simulations resolving the optimal scales to study complexity, i.e. $\Delta x = 100 \rm ~kpc/h \sim 142 \rm kpc$ for all simulations.

\begin{figure*}
\includegraphics[width=0.95\textwidth]{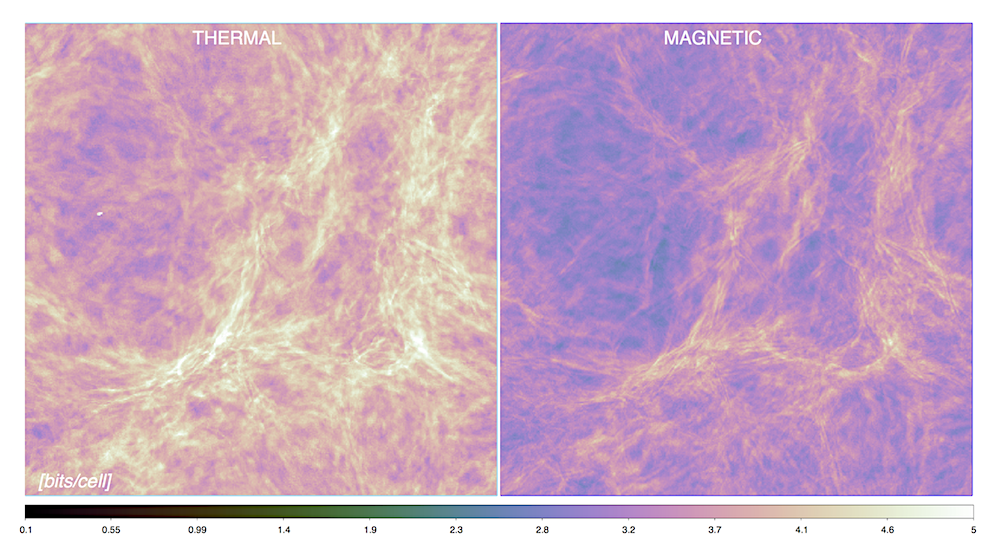}
\includegraphics[width=0.95\textwidth]{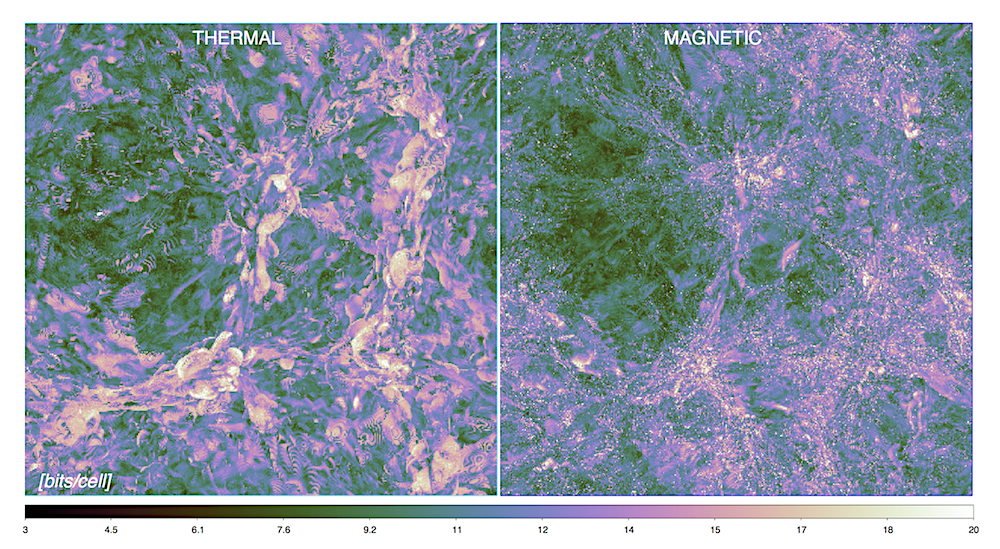}
\caption{3-dimensional renderings of the complexity of thermal energy (left) or of magnetic energy (right) across the full simulated volume. The top panels show the volume-weighted average complexity along the line of sight while the bottom panels give the maximum statistical complexity along the line of sight.}
\label{fig:3d}
\end{figure*}

\begin{figure*}
\includegraphics[width=0.95\textwidth]{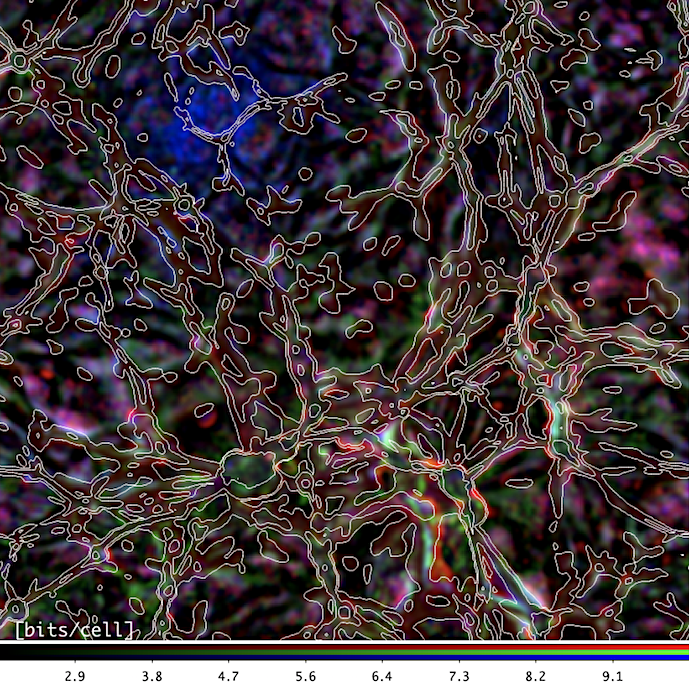}
\caption{Composite RGB image of complexity (red=thermal complexity, blue=kinetic complexity and green=magnetic complexity) for the same selection of Fig.\ref{fig:cut_fields}. The additional white contours give the distribution of thermal energy. }
\label{fig:rgb}
\end{figure*}

\begin{figure*}
\includegraphics[width=0.33\textwidth]{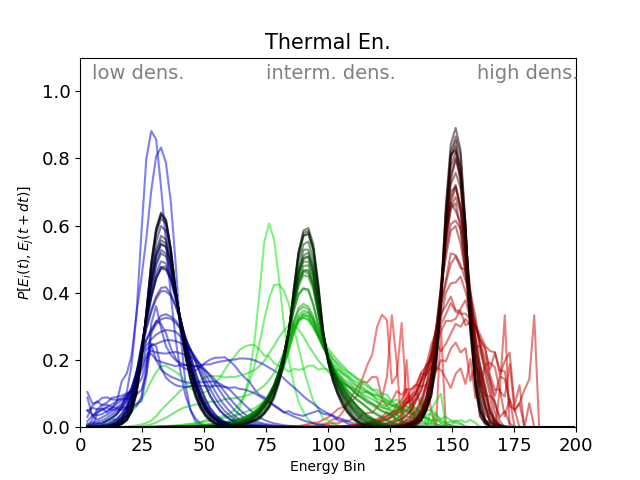}
\includegraphics[width=0.33\textwidth]{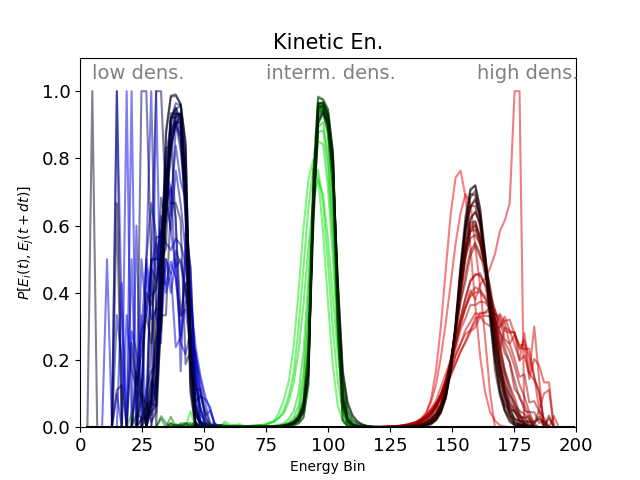}
\includegraphics[width=0.33\textwidth]{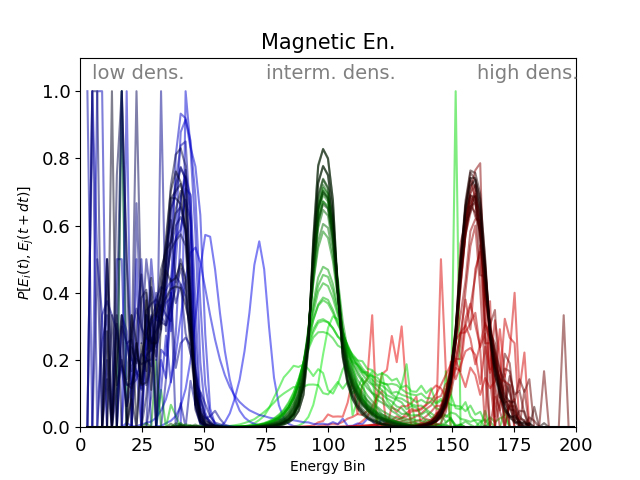}
\caption{Transition probability distribution for thermal, kinetic and magnetic energy (from left to right) as a function of time, for three reference energy bins marking low (blue colors), intermediate (green) and high (red) density environments, for run Cosmo0. The evolution of each probability distribution uniformly goes from the black line ($\approx 0.4$ Gyr since the begin of the simulation) to the lightest colors in each distribution ($\approx 13.7$ Gyr).}
\label{fig:pdf}
\end{figure*}

\begin{figure*}
\includegraphics[width=0.245\textwidth]{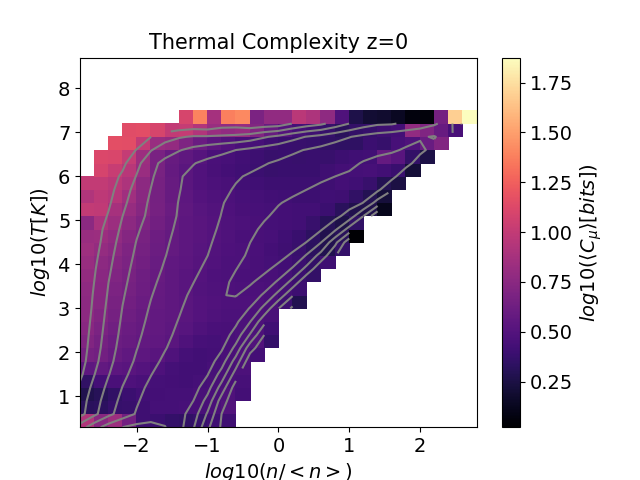}
\includegraphics[width=0.245\textwidth]{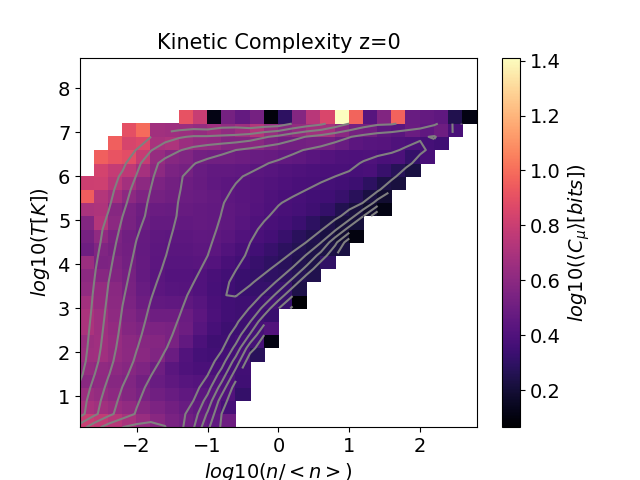}
\includegraphics[width=0.245\textwidth]{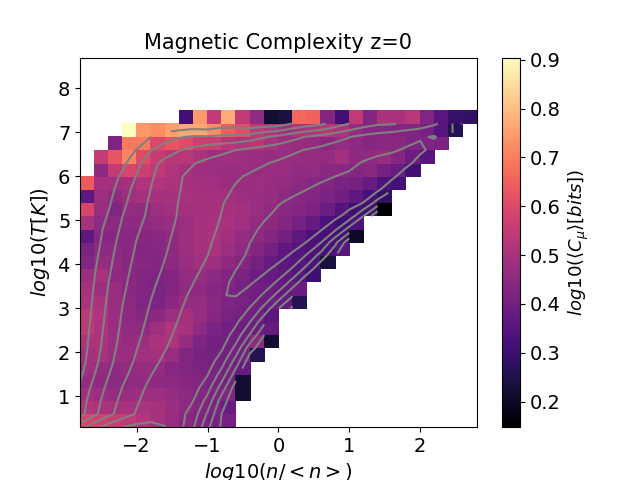}
\includegraphics[width=0.245\textwidth]{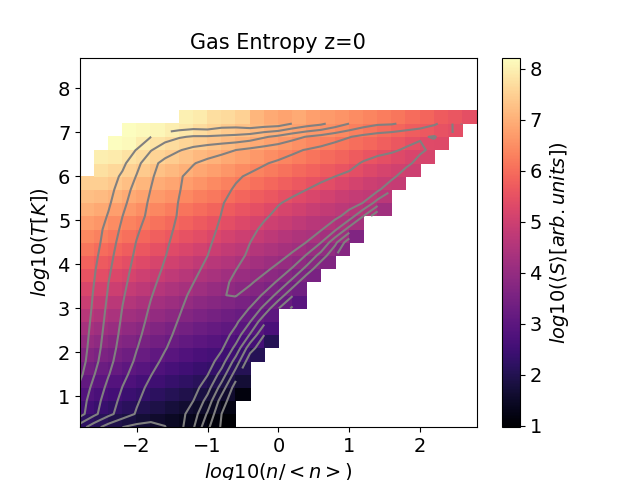}
\includegraphics[width=0.245\textwidth]{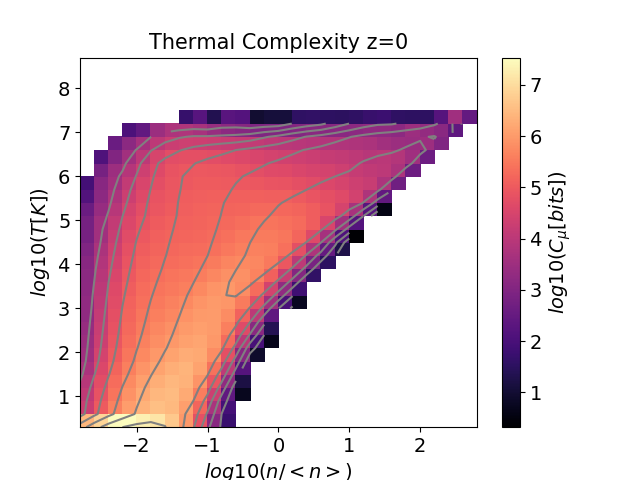}
\includegraphics[width=0.245\textwidth]{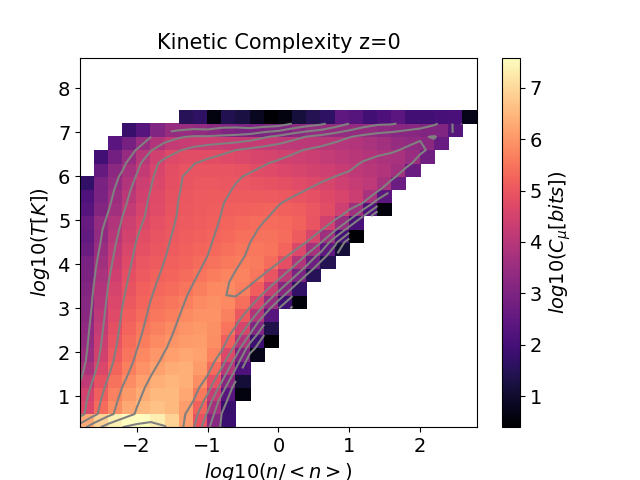}
\includegraphics[width=0.245\textwidth]{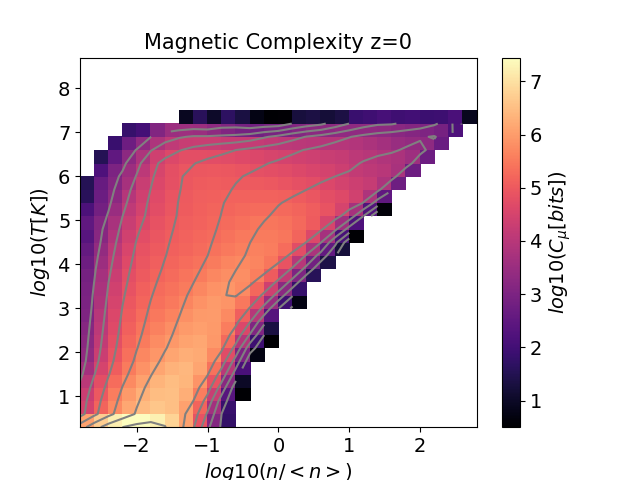}
\includegraphics[width=0.245\textwidth]{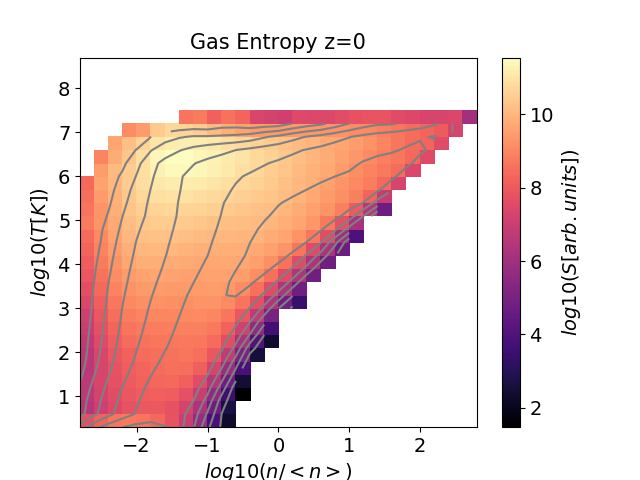}
\caption{Phase diagrams showing the average (top panels) or total (lower panels) complexity of the thermal, kinetic and magnetic energy at $z=0$ for the baseline model. The last column shows the average and the total  gas entropy for the same simulation and epoch. The grey contours in each panel show the gas density distribution (with logarithmic spacing of contours).}
\label{fig:phase1}
\end{figure*}

\subsection{Measuring information and complexity in a simulation}
\label{info}

While  the minimal information needed to {\it perfectly} describe a system is given by the {\it algorithmic complexity} \citep[e.g.][]{kolm,1995chao.dyn..9014C}, a more useful definition considered here and in Paper I, for the sake of practical applications, is the {\it statistical complexity} \citep[e.g.][]{1989PhRvL..63..105C,1998PhLA..238..244F,adami,prokopenko2009information}.

With the use of the statistical complexity it is possible to quantify how complex a particular realization of a given process is, independently of its "atomic"  distribution. 
In the case of simulations, this makes it possible to quantify the complexity of different realization of different sets of random initial conditions (or slightly different initial conditions) regardless of the byte-wise values of the final outputs. 

When analyzed in terms of statistical complexity, a purely random process is not complex as its entire evolution is explained by a simple statistics.  For the same reason, a perfectly periodic system also has a low level of statistical complexity as knowing how to compute the range of its statistical states over one period fully captures its future evolution \citep[e.g.][]{prokopenko2009information}.

Cosmological simulations evolve physical fields  (in a nutshell: gas and dark matter mass, their 3-dimensional velocity components, the internal gas temperature/energy and optionally 3-dimensional magnetic fields, as in this case, or any other physical field of interest) from early cosmological epochs until a late redshift. In Eulerian grid simulations explored here, a fixed comoving volume is partitioned into cells, and  the fluxes of matter, momentum and energy across the cells'interface determine the evolution of physical fields across the volume. 
The statistical complexity works by monitoring the transition probabilites across internal (discrete) states in which a system can be partitioned, and computing the Shannon entropy \citep[][]{1949IEEEP..37...10S} associated to each transition, directly from the recorded datastream of the simulation.  The statistical complexity also measures  how likely it is that a physical system simultaneously operates different tasks, a behaviour that makes its evolution increasingly harder to predict. Once a system is partitioned into sub-domains (e.g. cells in a mesh) or sub-levels (e.g. energy levels), the evolutionary track of each sub-element can be tracked over time, searching for patterns.  If a certain sub-element is found to always give the same output, its evolution is fully prescribed, and the overall complexity of the system lowers. \\

Very interestingly, the complexity analysis requires no knowledge about the physical laws involved in the evolution of system to analyze: it can be used to deduce the pattern of transitions or exchange between the internal simulation states as a function of time, regardless of the exact physical meaning of such transitions, and entirely based on the degree of freedom involved in the evolution of a specific computing element (i.e. a cell in a grid) in the simulation. The physical interpretation of such complexity can be than used a-posteriori to understand why the same set of physical laws can lead to complex or simple evolutionary patterns as a function of the underlying cosmic environment, or epoch. \\

Following \citet{va17info}, I monitored the evolution of cosmic matter in the simulated universe by focusing on the evolution of the energy fields associated to normal baryons: the gas kinetic energy ($E_k=\int (1/2) \rho v^2  dV$ within each cell), the gas thermal energy ($E_{t}=\int (3/2) k_b \rho T/(\mu m_p) dV$) and the magnetic field energy ($E_b=\int B^2/(8\pi) dV$), where the volume integral is over the cell size {\footnote{I did not explicitly consider the evolution of the dark matter distribution, considering that in all cosmological models in the ($\Lambda$)CDM framework the evolution of gas and of the dark matter components on $\geq 10^2 \rm ~kpc$ scales are closely coupled, hence the evolution of complexity in the gas component also gives the evolution of complexity in the dark matter distribution, within a small bias factor of order unity.}}. 

As already noted in \citet{va17info} there is no unique way of partitioning the internal energy of a cosmological simulation. 
For this reason, the arbitrary choice of how to partition the system into a set of discrete states is the result of a compromise between the need of keeping the  computing resources under control (as the computation of the statistical complexity scales as $\propto N_{\rm bin}^2$ (where $N_{\rm bin}$ is the number of energy bins) and the capability of resolving all relevant energy transitions between close timesteps.

The  statistical complexity is thus measured, at each timestep, in two basic steps:

\begin{itemize}

\item  first, by partitioning the simulation into discrete levels ($E_i$, with $1 \leq i \leq N_{\rm bin}$, $N_{\rm bin}$ being the total number of levels in the partition)
\item second, by measuring the frequency of transitions between a discrete energy level,  $E_i(t)$ at a given time $t$ transition to another level at a following epoch $t +\Delta t$, $E_j(t+\Delta t)$. 
\item third, by computing the transition probability distribution $P[E_i(t)|E_j(t+\Delta t)]$, directly from the $N_{\rm bin} \times N_{\rm bin}$ matrix encoding the distribution of  all possible transitions across energy levels. 

\end{itemize}

In the {\it fiducial} procedure, I employed $N_{\rm bin}=200$ equal energy bins in logarithmic space, ranging from the maximum and the minimum of each energy field, respectively, and consider a time spacing of $dt=5$ timesteps ($\approx  200 ~\rm Myr$) between snapshots. Variations of this setup are explored in Sec.\ref{variations}. 

The Shannon entropy associated to the probability of transition for each cell in the volume, with given 3-dimensional location $xyz$, is the 3-dimensional statistical complexity:

\begin{equation}
C_{\mu,xyz} = - P_{xyz} \log_2 P_{xyz},
\label{eq:complex}
\end{equation}

which is measured in [bits]. 

Each cell of our computing domain is therefore regarded as a processing unit, which is responsible for the production of a stream of $L$ symbols (where $L$ is the total number of epochs/timesteps) drawn from a ``vocabulary'' of $N_{\rm bin}$ words (i.e. energy levels). 

The matrix of transition probabilities is thus fully derived from the statistics of the datastream generated by the simulation, without referring to the underlying physics {\footnote{Variations of the above baseline metrics for statistical complexity have been also proposed in the literature, to incorporate effects of disequilibrium and extensivity \citep[e.g.][]{1995PhLA..209..321L,1998PhLA..238..244F}.}}.\\

An example of the transition probability matrix of our baseline simulation at $z=0.0$ is shown in Fig.\ref{fig:prob}. In the probability matrix, the diagonal 1-to-1 relation would correspond to the little complex case in which every $E_i(t)$ state 
is mapped to $E_i(t+dt)$, i.e. the energy states do not evolve. Conversely, 
the amount of spread around the diagonal line correlates with the complexity of the energy transitions in the datastream.

In order to display how the transition probability distribution evolves in time in the different environments, I present in Fig.\ref{fig:pdf} the sampling of it for three reference energy bins: a low energy one which roughly correspond to the low density regime of voids, the median one associated with 
matter sheets and cosmic filaments, and the high one associated with halos (e.g. Fig.\ref{fig:cut_fields}). The broadening of each distribution as a function of time leads to an increase of the statistical complexity as it measures the presence of large transitions between states in the simulation.  
In the case of thermal energy, the intermediate energy levels are characterized by a large spread in the probability distribution, which widens to high energy. This follows from the fact that the most radical changes in the thermal energy of cosmic gas typically happens in the low energy range, where $T \sim 10^4 \rm ~K$ gas is violently shock-heated by $\mathcal{M} \gg 10$ shocks to higher temperatures. The same class of shocks do not cause equally dramatic changes in kinetic energy (as follows from standard jump conditions), while the probability matrix of kinetic energy shows a larger spread in the high energy end. 
Finally, the magnetic energy shows significant spread in all environments. At the lowest magnetisation level, we can expect {\it numerical effects} to play a role, in the sense that a number of spurious fluctuations in the magnetic field are  expected in voids, in which the code also employs a (non-conservative) "dual energy formalism" to deal with hypersonic flows \citep[e.g.][]{enzo14}. A spread in $P_{\rm xyz}$ in the intermediate energy regime is instead more likely to be associated with the physical effect of magnetic field amplification (displayed by the fact that the green distribution gets increasingly skewed towards larger values of magnetic energy), which is a combination of gas compression and dynamo amplification, albeit at a relatively lower rate here, due to the coarse numerical resolution.

To perform the analysis of complexity, I saved $N_{\rm step} \sim 320-400$ snapshots for each run (the exact number depending on the number of timesteps that {\enzo} generated as a function of the Courant conditions of each run), sampling the cosmic evolution with a  typical time spacing of $\approx 50$ Myr among snapshots (i.e., smaller that the actual time difference used to computed complexity,  $\sim 200$ Myr in the fiducial procedure). 
It shall be noticed that  the computing problem here is non trivial, as the computation of statistical complexity for $N_E=3$ energy fields requires to generate an $N_{\rm bin} \times N_{\rm bin}$ matrix of transition probabilities across 2 subsequent snapshots for the entire sequence of $N_{\rm step}$ snapshots, for $N_E \times  n^3$ dataset at every iteration (where $n=400$ cells in this case). 
Compared to \citet{va17info}, in which routines in IDL were developed to compute statistical complexity and block entropy of cosmological simulations, in this work I implemented the above algorithm in  Julia language (https://julialang.org), for computing speed reasons.

A sample serial routine written in Julia (v.0.6.4)  to efficiently compute the statistical complexity in a set  of unigrid {\enzo}-HDF5 files is available at this URL{\footnote{https://github.com/FrancoVazza/JULIA/tree/master/INFORMATION}}. For a $400^3$ grid distribution, the computing time is $\approx 590 \rm ~s$ on an Intel Xeon(R) CPU E5-2620 core. \\

In \citet{va17info} I also presented applications of the "block entropy"  (i.e. the Shannon entropy of the entire sequence of "symbols" produced by processing units as a function of time,  \citealt[e.g.][]{Larson20111592,feldman1997bii,Crutchfield03}) and of the "efficiency of prediction" (i.e. the ratio between statistical complexity and the time difference of the block entropy, which marks the scale(s) at which  making predictions of the future evolution of the system is more efficient,
\citealt[][]{2004PhRvL..93n9902S,prokopenko2009information}). These proxies of complexity are numerically even more demanding as they require to process the entire time sequence of symbols in the simulation, which is only doable for small sub-selections in the entire domain, and will be subject of future investigations with more efficient numerical algorithms.

\begin{figure*}

\includegraphics[width=0.245\textwidth]{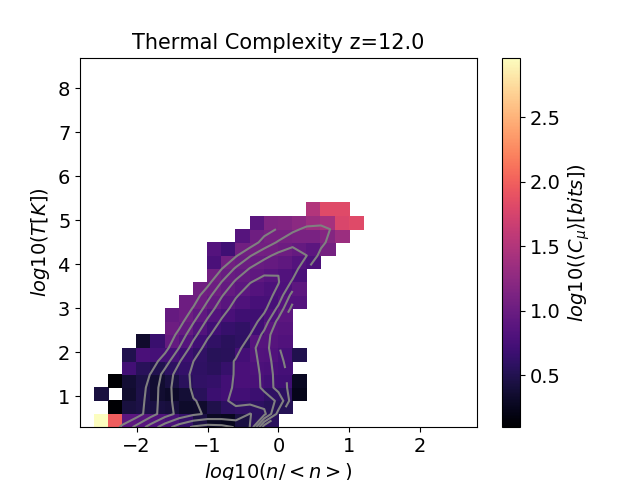}
\includegraphics[width=0.245\textwidth]{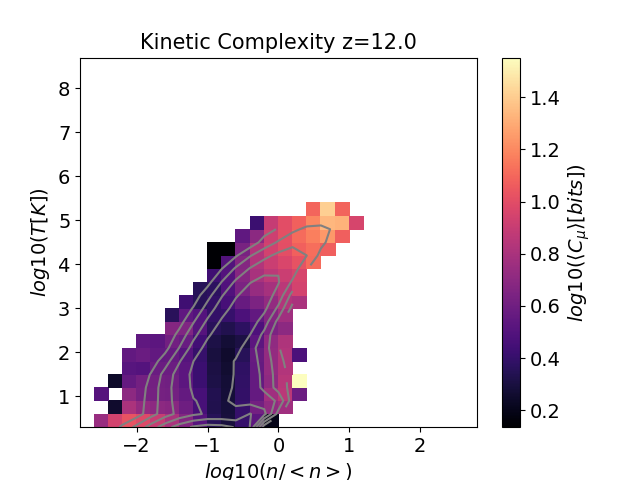}
\includegraphics[width=0.245\textwidth]{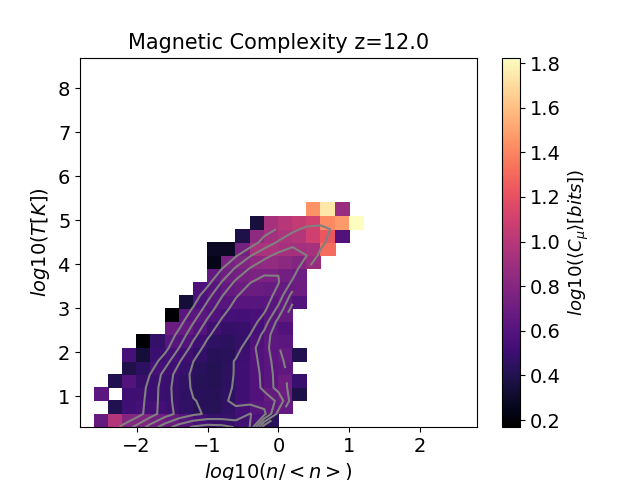}
\includegraphics[width=0.245\textwidth]{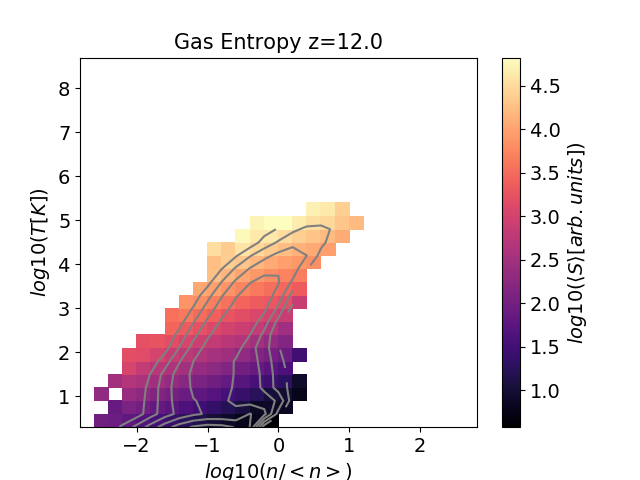}
\includegraphics[width=0.245\textwidth]{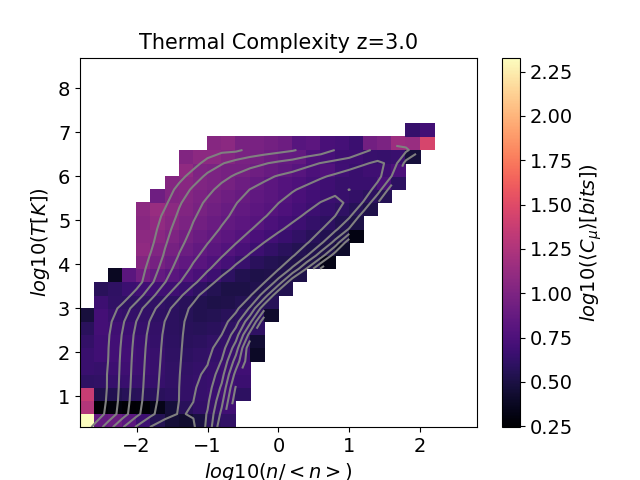}
\includegraphics[width=0.245\textwidth]{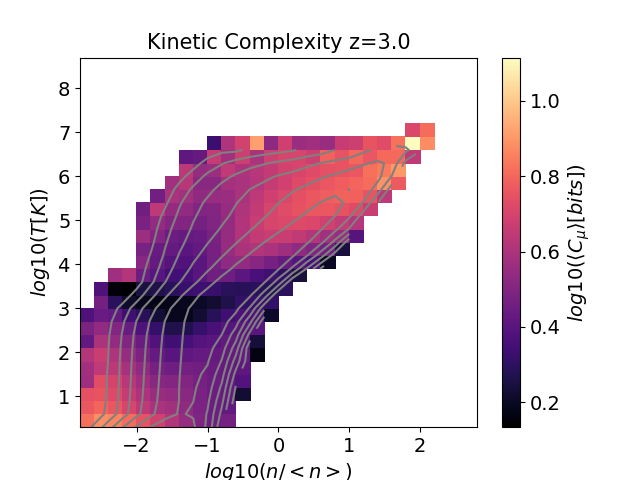}
\includegraphics[width=0.245\textwidth]{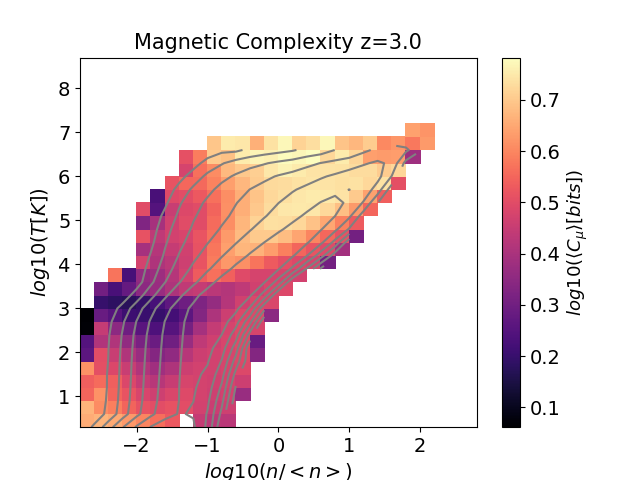}
\includegraphics[width=0.245\textwidth]{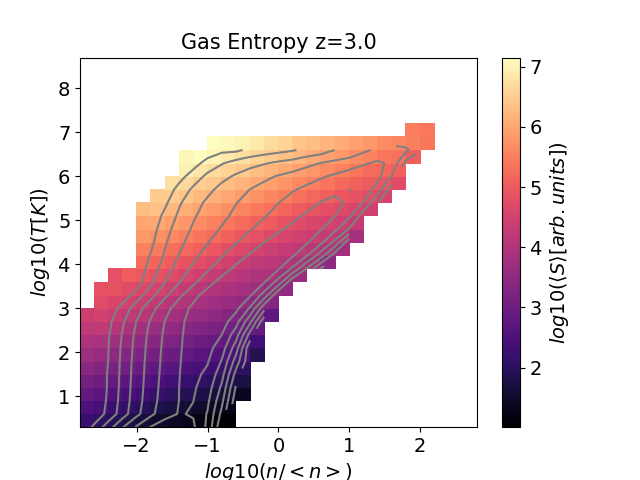}
\includegraphics[width=0.245\textwidth]{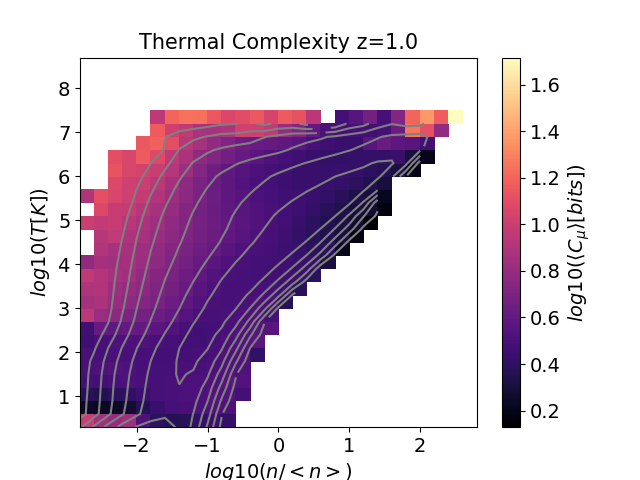}
\includegraphics[width=0.245\textwidth]{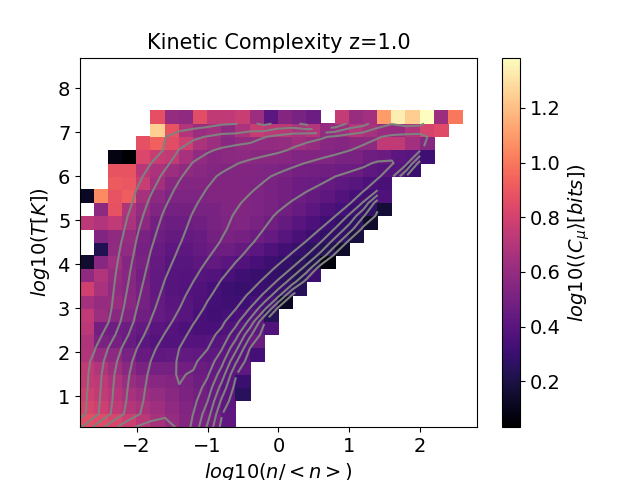}
\includegraphics[width=0.245\textwidth]{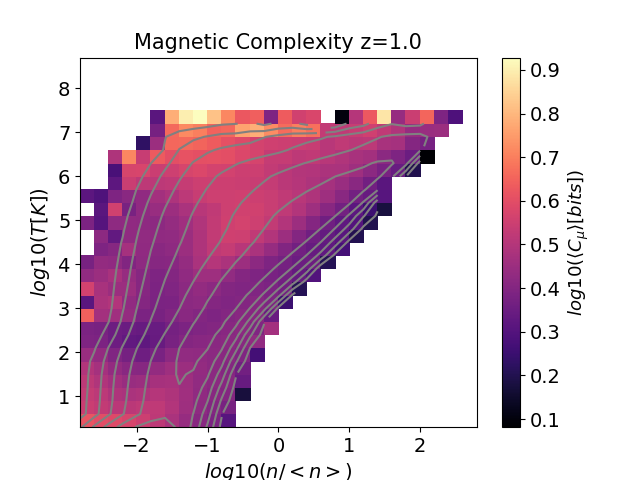}
\includegraphics[width=0.245\textwidth]{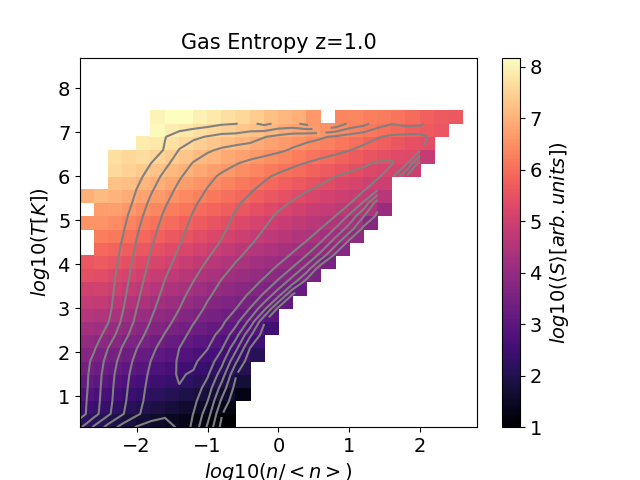}
\caption{Phase diagrams showing the average complexity in the thermal, kinetic and magnetic energy (first three columns), as well as the average gas entropy (last column) in our baseline simulation at $z=12$, $z=3.0$ and $z=1.0$.}
\label{fig:phase2}
\end{figure*}

\section{Results}
\label{sec:res}

\subsection{What is the complexity of the cosmic web today?}
\label{z0}
First, I focus on the global view of statistical complexity in the cosmic volume at $z=0$ in the baseline model. 

Unlike many other physical quantities of interest, the 3-dimensional distribution of complexity in the simulated Universe has never been imaged before in the literature, hence I give here a few different views here to best give a sense of its morphology: 
Figures \ref{fig:cut_fields}-\ref{fig:cut_info200} give a 2-dimensional view of the distribution of the energy fields and of their measured $C_{\rm \mu}$ for a thin ($100$ kpc/h) slice crossing the centre of the simulated box;  Figure \ref{fig:3d} shows the average and maximum complexity along one line of sight through the entire volume (limited to the thermal and magnetic complexity) while  Fig.\ref{fig:rgb} gives the composite red-green-blue (RGB) image of complexity in the simulated volume, assigning  the complexity of each energy field to a different color channel. \\

The distribution of all energy fields closely matches the one of gas density and dark matter (not shown) and  traces the filamentary structure of the cosmic web, with maxima corresponding to the peaks of the  energy distribution, which are the self-gravitating matter halos in the volume. 
The visual inspection shows that the 
volume filling factor of the different energy fields is not the same: the process of virialization irreversibly transfers energy from the extended distribution of kinetic energy into more spatially confined distributions of  thermal gas energy and of magnetic energy, through the combination of simple compression and irreversible shock heating and small-scale dynamo amplification, respectively (albeit in this case with a significant quenching of the process due to the limited numerical resolution). 

The different views of complexity in the same volume show that this has a broader spatial distribution compared to the one of the energy fields. This means that regions with a very different energy ratio may undergo an equally complex evolution at a given time. 
The projected view along the full $40 ~\rm Mpc/h$ line of sight best help visualizing the 3-dimensional structure of the complexity maxima in thermal energy, which are closely tracing shocks surrounding filaments and massive halos in the box: this is exactly where  the conversion of kinetic energy into thermal and magnetic energy occurs, at a rate that depends on the local flow. 
The largest jumps in the internal gas energy in the simulation are indeed found in the presence of strong shocks, as $E_{\rm th,2}/E_{\rm th,1} \propto \mathcal{M}^2$. In simulated structure formation, such strong shocks are always found at the periphery of structures, and mark the first episode of strong (irreversible) heating of gas with a pre-shock temperature of $T \sim 10^3-10^4 ~\rm K$ \citep[e.g.][]{ry03}. On the other hand, strong shocks can only produce a mild jump in magnetic energy even in the limit of very large Mach number, $E_{\rm B,2}/E_{\rm B,1} \approx 4^{4/3}=\rm const$,  while the random twisting of magnetic field lines by the turbulence developed in halos \citep[e.g.][]{2019MNRAS.486..623D} overall results into a sharp increase of magnetic complexity in the densest environment, leading to values which are of the same order of the other two fields.

Even if the energy conversion at shocks is very simple physics, predicting which regions in the simulations will undergo shock heating in the next timestep(s) is a more {\it complex} question, involving the analysis of the local thermodynamic conditions in the 3-dimensional neighborhood of each cell \citep[e.g.][]{ry03,va11comparison}: a computing task that ultimately requires an extra amount of information, compared to what is necessary to describe the evoluton of smoother regions of the cosmic web.

The vast majority of the outer infall regions of galaxy clusters and filaments is still undergoing gas and dark matter accretion (either in the form of smooth or clumpy accretions), and each single accretion event can radically change any pre-existing level, by dissipating a large fraction of the kinetic energy developed during the infall. 
Therefore, the periphery of structures in the cosmic web is always
 characterized by high level of complexity, of order $C_{\rm \mu} \geq 10-10^2$ bytes/cell.\\
 
On the other hand, the internal volume of halos is in general less complex than their outermost regions,  because single perturbations on a short timescale can hardly change pre-existing energy levels by a large amount. Moreover, because the internal
regions of clusters (and, to a lesser extent, of filaments) underwent virialization during their formation epoch, gas perturbations are always sub-dominant compared to the cluster energy, leading to weaker shocks ($\mathcal {M} \leq 5$, e.g. \citealt{ry03}) and to sub-sonic turbulent velocity fluctuations \citep[e.g.][]{mi14,sch16}. An exception is represented by magnetic energy, 
whose evolution within clusters can be made significant complex by the small-scale dynamo amplification promoted by turbulence \citep[e.g.][]{review_dynamo}. This is also visible by the dominance of "green" colors within halos in the RGB superposition of complexities in Fig.\ref{fig:rgb}.  

Finally, the cell-wise complexity tends to be much lower in cosmic voids,  owing to the relatively simple evolution there, which is mostly ruled by adiabatic expansion. \\

The distribution of statistical complexity as a function of cosmic environment is well rendered by the phase diagrams in Fig.\ref{fig:phase1}, in which I show the average and the total  statistical complexity for each energy form. The last column also gives the  total and average gas entropy ($S \propto T/n^{2/3}$) for the same simulation. 

The most complex gas phase in the simulated volume is indeed found in the high temperature range ($\sim 10^4 - 10^7 \rm ~K$) and across a wide range of densities, with $C_{\rm \mu} \sim 10-10^2 \rm ~bits/cell$. Lower values ($C_{\rm \mu}\sim 10 \rm ~bits/cell$) are found 
at high density $n/\langle n \rangle \geq 10-10^2 $ and are correlated with halos in the volume.  Finally, the low density and low temperature environment of voids is characterized by low complexity ($C_{\rm \mu} \leq 2-5 ~\rm bits/cell$).  \\

The distribution of gas entropy  in the phase diagram 
shows an overall similar distribution to the one of complexity, which is not surprising because entropy in this simple non-radiative simulation is mostly increased by the irreversible action of shocks.

Moreover,  the correspondence between gas entropy and statistical complexity is expected from the basic definition of entropy in statistical thermodynamics:

\begin{equation}
    S=-k_B \sum P_i \log P_i
    \end{equation}
  (where $P_i \propto e^{\frac{\epsilon_i}{k_B T}}$ is the probability of an energy state $\epsilon_i$ at temperature $T$, in the Boltzmann distribution), which is very similar to the definition of statistical complexity in Eq.\ref{eq:complex}.\\

When the total complexity within each gas phase is integrated over the entire distribution of cells in the volume (see lower panels of Fig.\ref{fig:phase1}), the relative contribution from the different cosmic environments is more biased towards the low density part of the cosmic web: this is understood because voids contribute to $\sim 90\% $of the volume of the Universe, vs $\sim 9\%$ of filaments and $\leq 1\%$of halos \citep[e.g.][]{iapichino11,2014MNRAS.441.2923C}.

\begin{figure}
\includegraphics[width=0.45\textwidth]{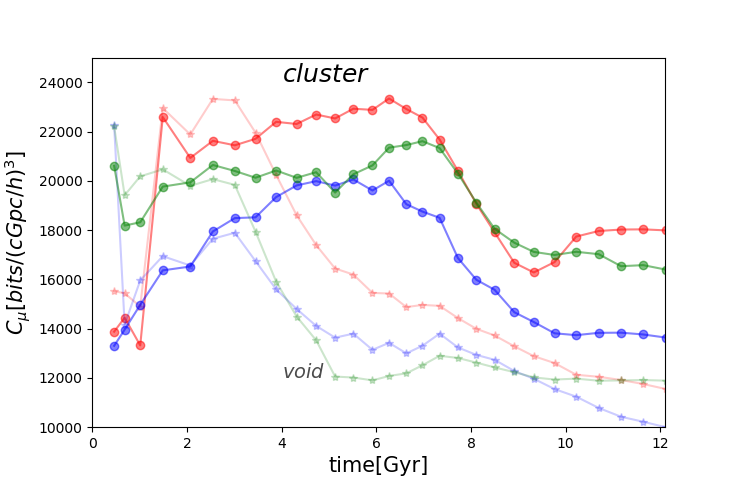}
\caption{Evolution of complexity for sub-selection centred on a massive halos or on a empty region in the simulation, in both cases a fixed comoving $\rm 3^3 Mpc/h^3$ volume. }
\label{fig:evol_env0}
\end{figure}

\begin{figure}
\includegraphics[width=0.45\textwidth]{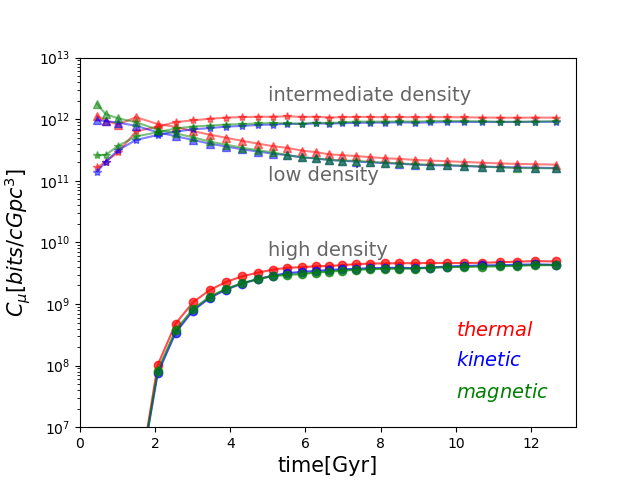}
\caption{Evolution of the total complexity in the baseline run, for different overdensity selections of cells in the cosmic volume  (see Sec.3.2).}
\label{fig:evol2}
\end{figure}

\begin{figure}
\includegraphics[width=0.45\textwidth]{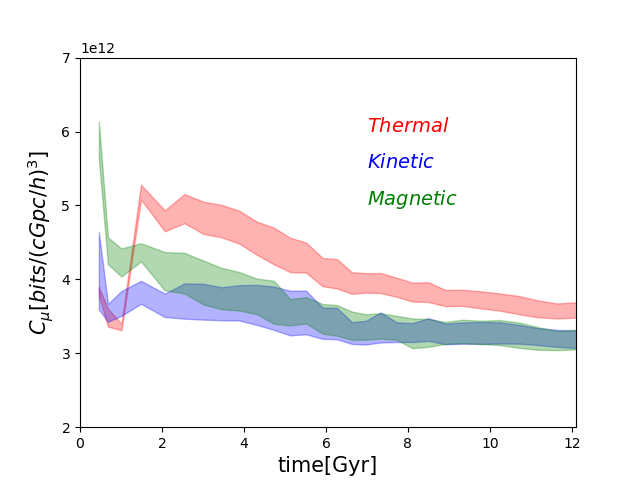}
\caption{Effects of cosmic variance on the evolution of complexity, measured in four different sub-boxes of $15^3 \rm ~Mpc/h^3$ within the main simulation.}
\label{fig:var}
\end{figure}

\begin{figure*}
\includegraphics[width=0.33\textwidth]{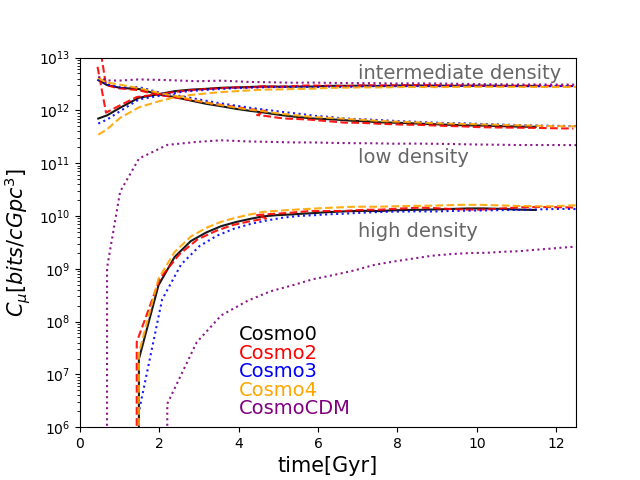}
\includegraphics[width=0.33\textwidth]{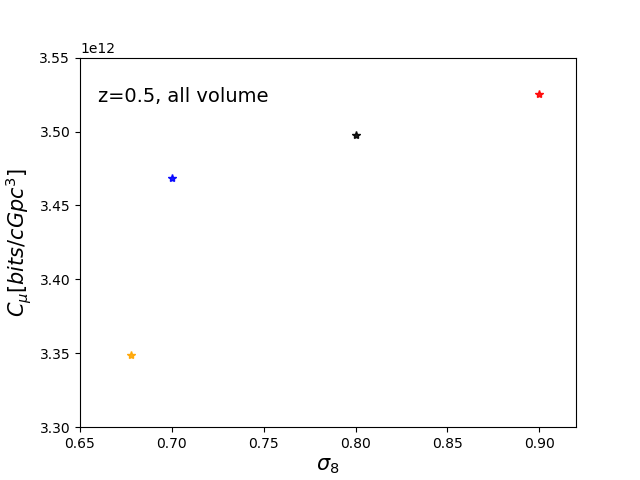}
\includegraphics[width=0.33\textwidth]{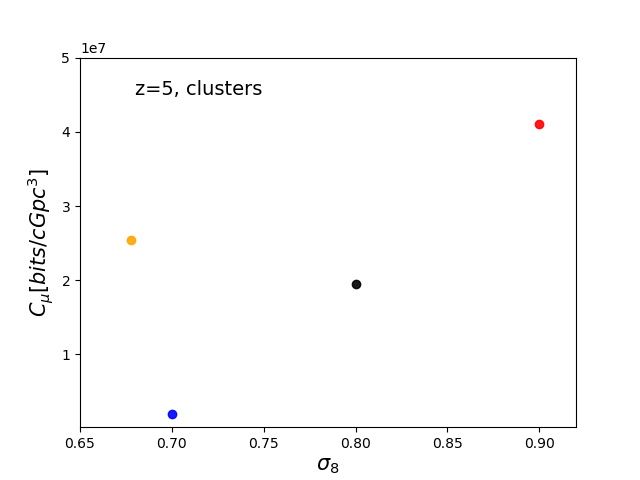}
\caption{First panel: comparison of the cosmic complexity (here the total of thermal, kinetic and magnetic complexity) for variations of the assumed cosmological model (see Tab.1 for details). Second and third panel: dependence of complexity on the $\sigma_8$ of simulations, considering the total complexity in the volume at $z \approx 0.5$ (centre) or the complexity of halos at $z \approx 5$.}
\label{fig:evol_cfr}
\end{figure*}

\begin{figure}
\includegraphics[width=0.45\textwidth]{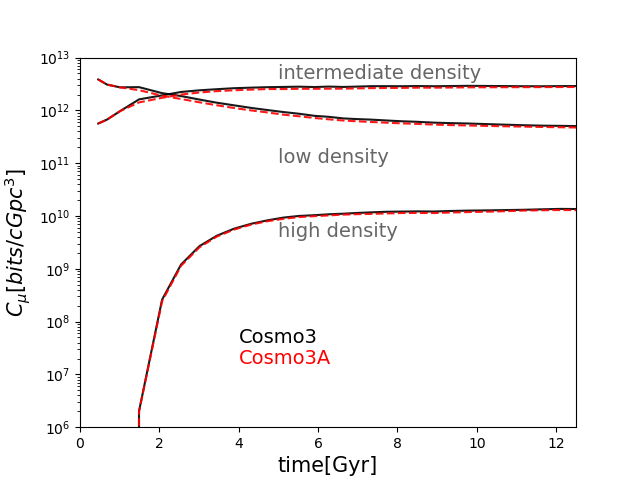}
\caption{Total cosmic complexity for run Cosmo3 and run Cosmo3A (with identical cosmological parameters but a different random realization of the initial phases for the matter density and velocity perturbations), as a test for cosmic variance on large scales. }
\label{fig:evol_cfr_var}
\end{figure}

\subsection{When did complexity emerge?}
\label{evolution}

The emergence of complexity as a function of time can be investigated by applying the same algorithm of Sec.\ref{info} to all available snapshots in the past of the baseline Cosmo0 model. 

The phase diagrams of  Fig. \ref{fig:phase2} show the evolution of the average statistical complexity for the three energy fields, as well as the evolution of the average gas entropy, for our baseline model at $z=12$, $z=3.0$ and $z=1.0$. Interestingly and opposite to what seen in the  $z=0$ phase diagram (Fig.\ref{fig:phase1}) in remote cosmic epochs the most complex environment was the highest density one, associated at all epochs with the formation of the first structures via gravitational collapse of the primordial density fluctuations.  
The process is always followed by the 
virializaiton of the infall kinetic energy acquired by gas during the collapse, which is again mediated by shocks (see similar patterns in the entropy phased diagram in the last column). \\

In the very early Universe (e.g. $z=12$ in the figure, $\approx 0.37$ Gyr since the start of the simulation) the gas temperature did not exceed $\sim 10^5 \rm ~K$ , as prior to virialization within halos the kinetic energy is overall the dominant energy form at all scales. 

Later on and down to $z \sim 1$, the rapid fluctuations of kinetic energy dominates the statistical complexity in the density regime typical of halos, while the irreversible heating episodes add increasingly more complexity to the low density and high temperature outskirts of clusters and filaments (where also gas entropy reaches its maximum).  As the evolution of magnetic fields is tightly coupled to the one of kinetic energy (via the induction equation in the ideal MHD case considered here) also the evolution of distribution of magnetic complexity as a function of gas phase is quite similar to the kinetic case. 

As the simulation proceeds, the maxima of complexity drift to a lower density environment, and by $z \sim 1.0$ the distribution is quite similar to what is also measured at $z=0$, again similar to the distribution of gas entropy (Fig.\ref{fig:phase1}).
Therefore, while the densest structures continue to accrete matter, expand and tend to a virial equilibrium in their internal regions, to a first approximation their complexity budget is already in place  at the epoch of $\sim 5 ~ \rm Gyr$.
\\

Therefore, the formation site of halos in the simulation is where complexity first emerges in the cosmic volume, to later diffuse out towards 
the lower density Universe, following the growth of the density perturbation that surrounds structures of the cosmic web. 
Figure \ref{fig:evol_env0} zooms onto the evolution of complexity in a comoving $3^3 \rm Mpc/h^3$ region centred on a $\sim 10^{14} M_{\odot}$ halo, and compares it with the evolution of complexity in an empty region within a void, for the baseline model. 
Clearly, the evolutionary tracks in the two environment are well detached already $\geq 2$ Gyr after the start of the simulation, i.e. during the   first collapse leading to the formation of the halo's progenitor.  From this point on, the complexity is all fields remained a factor $\sim 2$ larger in the case of the cluster region, and was subject to sporadic evolution following the dynamic history of the halo, which remained quite active until an epoch of $t \sim 9 \rm Gyr$ ($z \sim 0.4$). 
On the other hand, the complexity in the void region smoothly declines as a function of time as the evolution of all fields is governed by the (simpler) effect of adiabatic expansion. It shall be noticed that not all voids in the simulated volume evolve in a so simple way: for example, the emptiest voids in the simulation are associated with significant motions of gas in expansion towards the surrounding structures, which can produce extended patches of high kinetic complexity (e.g. see the "blue" excess of kinetic complexity in the top left corner of Fig.\ref{fig:rgb}). For this reason, quantifying the overall distribution of complexity across cosmic environment requires to average $C_{\rm \mu}$ for all different structures formed within different ranges in overdensity.

When  the complexity is integrated over the different matter phases  in the cosmic volume, most of "cosmic complexity" is overall contributed by the gas in the moderate overdensity typical of cosmic filaments. 
This is quantified in Figure \ref{fig:evol_env0}, in which  I show the evolution of the total complexity (normalized to a comoving $\rm Gpc^3$ volume) after 
dividing the contribution from the simulated volume into different gas overdensity regimes, which can be used to differentiate  "voids" (i.e. cells with $n/\langle n \rangle < 0.5$) from "sheets and/or filaments" ($0.5  \leq  n/\langle n \rangle  < 50$)  from "clusters" ($n/\langle n \rangle \geq 50$),  using fiducial values for a quick distinctions of cosmic web components \citep[e.g.][]{2016MNRAS.462..448G}. For simplicity, in the reminder of the paper I will refer to the above as"low", "intermediate" and "high" density environment, respectively.

All fields show a first peak of complexity for $\leq 1$ Gyr, as the clustering start from smooth linear initial conditions, hence multiple transition probabilities establish for the first time, according to where the gas resides. 
The large peak of complexity in the magnetic energy is explained by the artificial initialization of the seed magnetic field in the volume, which is for simplicity (and in order to enforce the $\nabla \cdot \vec{B}=0$ condition) set to spatially constant value at $z=30$. 
Already after the first Gyr  (i.e. for $z\leq 5-6$) the volume integrated complexity is dominated by the intermediate density gas phase, and this gas phase contributes to $\sim 90\%$ of the total complexity, while only $\leq 1\%$ is in total contributed by the densest phase of halos.
While the complexity associated with the underdense Universe decreases over time, the one in the intermediate energy regime grows fast during the first $\rm Gyr$ and then sets to an almost constant value.  Due to their late formation, self-gravitating halos in the volume form and become complex on average only at later redshift ($z \leq 2$). In all cases, the thermal complexity is found to be the dominant for the vast majority of the simulated cosmic evolution.
Normalized to a reference comoving $\rm Gpc^3$ volume, the total complexity of the cosmic web  is $C_{\rm \mu}\sim 10^{12} \rm ~bits$ at most epochs. \\

In order to assess the impact of cosmic variance across the volume, I computed in Fig.\ref{fig:var} the evolution of $C_{\rm \mu}$ for four independent cubic regions of  $15^3 \rm ~Mpc^3$ from the same baseline run. From the width  of the distribution of each complexity field it can be estimated that cosmic variance can typically result into a small $\Delta C_{\rm \mu} \leq 20 \%$ uncertainty at most epochs, consistent with the fact that most of the volume-integrated cosmic complexity is contributed by the low/intermediate density gas, whose variance on $\geq 10 \rm ~Mpc$ scales is overall rather small.

\begin{figure*}
\includegraphics[width=0.999\textwidth]{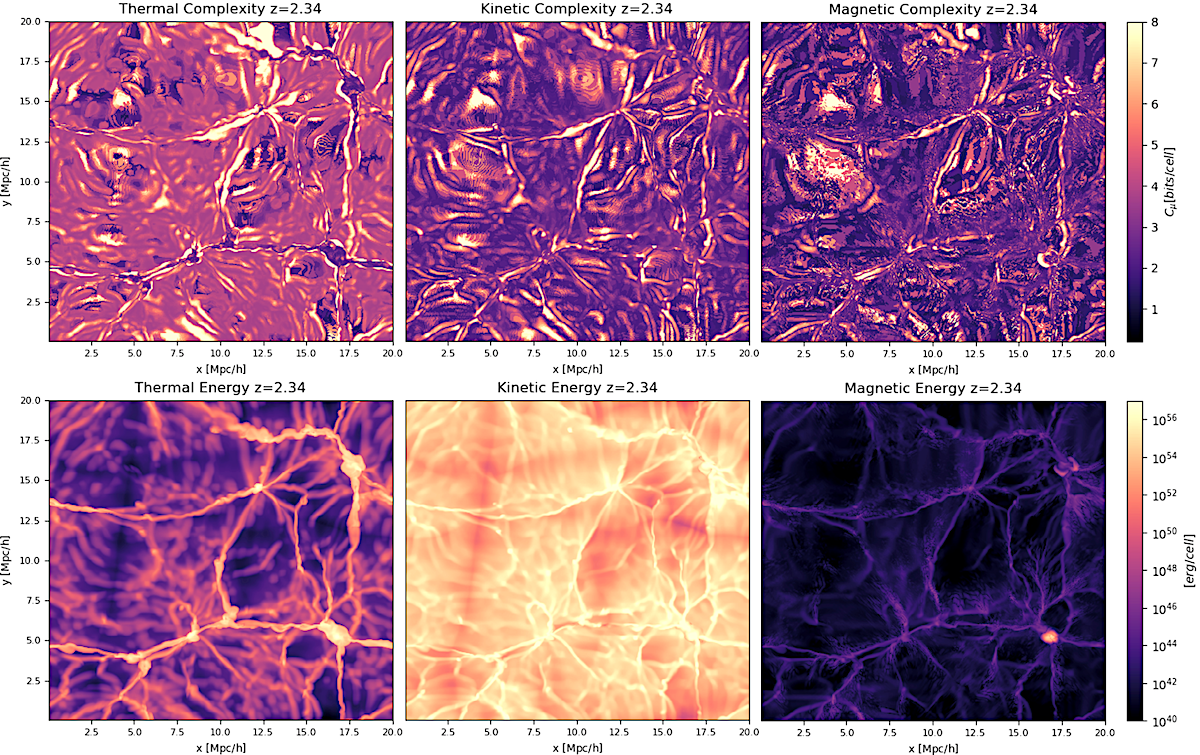}
\caption{Distribution of statistical complexity (top panels) and  energy (lower panels) for a thin slice with width $100$ kpc/h through the Cosmo0AGN run at $z=2.34$.}
\label{fig:cqh1}
\end{figure*}  

\begin{figure*}
\includegraphics[width=0.999\textwidth]{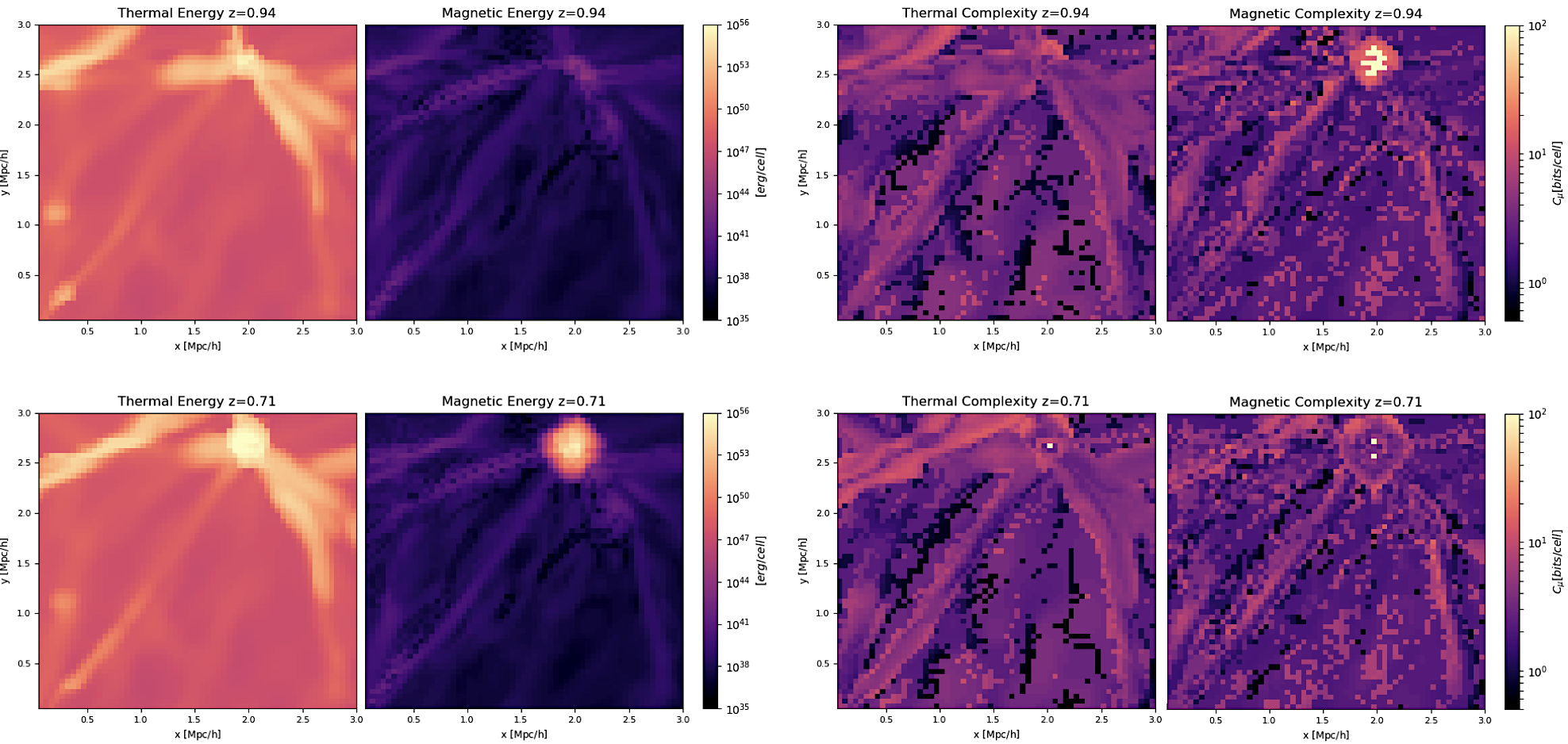}
\caption{Zoomed maps of thermal and magnetic energy (first two columns) and of thermal and magnetic complexity (last two columns) for run Cosmo0AGN2 (with radiative gas cooling and magnetic/thermal feedback from AGN) at $z=0.94$ (top row) and $z=0.71$ (bottom row), for a thin slice with width $100$ kpc/h.}
\label{fig:cqh2}
\end{figure*}

\begin{figure}
\includegraphics[width=0.45\textwidth]{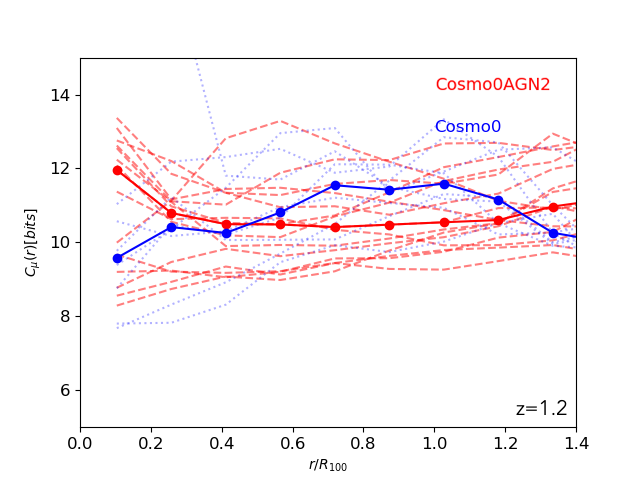}
\includegraphics[width=0.45\textwidth]{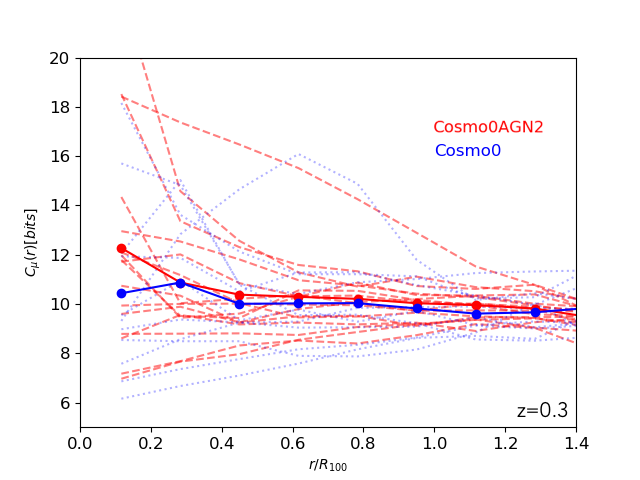}
\caption{Radial profiles of statistical complexity in $M_{\rm 100}\geq 10^{13} M_{\odot} $ halos in Cosmo0 and Como0AGN2 runs, for $z \approx 1.2$ and $z \approx 0.3$. The dotted/dashed lines show the radial profile of each object (normalized to $R_{\rm 100}$), while the solid lines show the median profile in the two runs.}
\label{fig:prof_cqh}
\end{figure}

\begin{figure}
\includegraphics[width=0.45\textwidth]{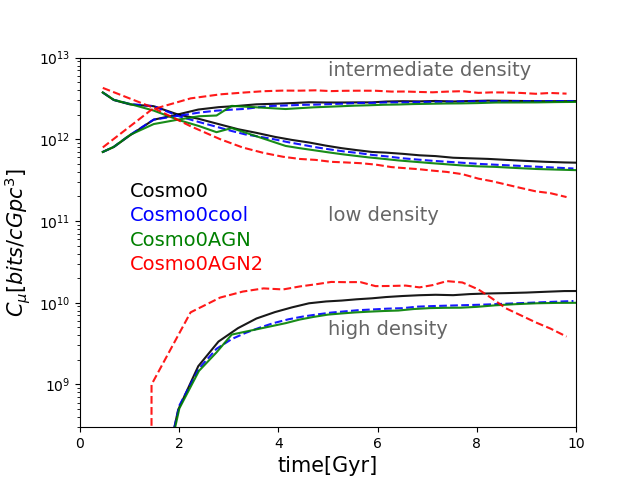}
\caption{Comparison of the total cosmic complexity for run Cosmo0, Cosmo0cool (with radiative cooling), Cosmo0AGN and  Cosmo0AGN2 (with cooling and AGN feedback) as a function of time. }
\label{fig:evol_cfr_cool}
\end{figure}

\section{Discussion}
\label{variations}

In this Section I will present additional test for the robustness of the above statistics, against a few numerical and physical variations with respect to the baseline methodology used in the previous Sections. 

\subsection{Physical variations: cosmology, cosmic variance and gas physics}

First, I tested the dependence of cosmology on the cosmic complexity comparing runs Cosmo2 and Cosmo3, Cosmo4 and CosmoCDM against the reference cosmological model (Cosmo0): the first two assume the same cosmology but a higher or lower
normalization for the initial spectrum ($\sigma_8=0.9$ and $0.7$, respectively), the second adopts the cosmological parameters of the PLANCK cosmology \citep[][]{2016A&A...594A..13P} while the last adopts a Cold Dark Matter cosmology without dark energy.
The first panel of Figure \ref{fig:evol_cfr} gives the evolution of the total integrated complexity in the above model variations.
The difference between all $\Lambda$CDM models is very small at all most epochs and for all energies. However, there is a small dependence between the initial amplitude of $\sigma_8$ and the complexity measured  a fixed epoch, as shown in the second and in the third panel of Fig.\ref{fig:evol_cfr}.
At low redshift, the total complexity of all fields is increased by a few percent if $\sigma_8$ is increased; the trend is also visible if the Cosmo4 run (with $\sigma_8=0.678$) is considered, even if also the rest of its cosmological parameters are different. 
Due to their different epoch of formation, also the complexity of halos at high redshift (third panel) shows evidence of a dependence on $\sigma_8$, albeit the correlation is later lost at lower redshifts.

The above trends can be understood considering that a larger value of $\sigma_8$ implies that the collapse of self-gravitating perturbations starts earlier in time, as well as that a higher $\sigma_8$ produces more matter substructures \citep[e.g.][]{2012ARA&A..50..353K}, which in turns promotes the emergence of complex patterns earlier in time, and a higher value of $\sigma_8$ also promotes a faster and more volume filling development of shocks in the cosmic volume \citep[][]{va09shocks}.

 Larger differences in the complexity of the low and high density environments are instead found when the CDM model is compared with $\Lambda$CDM cosmology, while the intermediate density environment remains equally complex. Such differences can be understood considering that the while the total number of clusters forming in the volume approximately scale as $\Omega_M \sigma_8^0.5$  \citep[e.g.][]{2002ARA&A..40..539R} and is therefore similar in all models, the growth of cluster is much delayed in the CDM case \citep[e.g.][]{2001ApJ...551...15B}. Hence complex
pattern driven by matter accretion in halos are only emerging later in time.

In conclusion, these tests have shown that small variations with respect to the assumed fiducial  $\Lambda$CDM cosmology  ($\Omega_M=0.3$, $\Omega_{\rm \Lambda}=0.7$, $\Omega_b=0.04$, $\sigma_8=0.8$) contribute to an uncertainty of the cosmic complexity at most epochs of order  $\Delta C_{\rm \mu}\leq 10\%$, with a small dependence on the $\sigma_8$ parameter.\\

 I have also tested the effect of cosmic variance by computing complexity in a parent resimulation of the Cosmo3 model, obtained with different  random phases for the initial density and velocity perturbation field (run Cosmo3A). Figure \ref{fig:evol_cfr_var} shows that when the complexity is integrated on $\gg 10^3 ~\rm Mpc^3$ volumes, the variance is extremely small, i.e. $\Delta C_{\rm \mu} \leq 4-5\%$ as a fractional difference between the two (entirely different) runs, for all environments and all epochs. This ensures once more that the view of complexity obtained in this relatively small simulated volume is representative of the cosmic average, and that  extrapolations onto larger scales (as in Sec.~\ref{sec:conclusions}) can be performed.\\

Lastly,  I have tested the impact of additional non-gravitational processes on cosmic complexity, by computing the statistical complexity for the Cosmo0cool run (which includes radiative gas cooling) and for two additional runs including thermal and magnetic feedback from AGN, testing two fixed energy per events ($10^{58}$ erg/event in Cosmo0AGN and $10^{59}$ erg/event in Cosmo0AGN2, respectively, see Sec.\ref{subsec:sim} for details). 
The panels in Fig.~\ref{fig:cqh1} show the distribution of complexity in a fraction of the simulated volume at $z=2.34$, as well as of the corresponding energy fields, for run Cosmo0AGN2.
In general, the large-scale distribution of complexity is very similar to what is found in the baseline (non-radiative) simulation, with the only exception of the neighbourhood of regions interested by AGN feedback, as in the lower right corner of the image.
A close-up view of a galaxy group affected by AGN is shown in Fig.~\ref{fig:cqh2}, before and after a powerful feedback event which injected magnetic fields and additional thermal energy in the core of this group. 
The corresponding maps of thermal and magnetic complexity clearly shows peaks of complexity where the AGN feedback episode took place {\footnote{It shall be noticed that, in order to preserve the $\nabla \cdot \vec{B}=0$ condition, the implementation of magnetic feedback adopted here is the one injecting magnetic field dipoles at the opposite sides of the cell used for the thermal AGN feedback, and therefore the peaks of complexity associated with feedback event ares by construction not exactly co-spatial.}}. The inclusion of AGN activity thus creates an additional mechanism for the sudden variation of magnetic and thermal energy values in the simulation (as well of kinetic energy via outflows, not shown here). This, in turn, adds complexity to the evolution of baryons, at least in the close proximity of AGN events, up to $C_{\rm \mu} \sim 10^2$ bits/cell (see last column of Fig.\ref{fig:cqh2}), as already noticed in earlier work  \citep{va17info}. 

A more systematic look to the impact of AGN feedback on the cluster population is given by the comparison of the average radial profiles of complexity given in Fig.\ref{fig:prof_cqh}, for all halos more massive than $10^{13} M_{\odot}$ in runs Cosmo0 and Como0AGN2 at high and low redshift ($z \approx 1.2$ and $z \approx 0.3$, respectively). In the proximity of cluster cores, the impact of AGN on complexity can be extremely large for single objects, while in general the median over the samples shows a $\sim 20-30 \%$ excess in the AGN case in the cluster core regions. It shall be noticed that this difference is smaller than what I earlier reported in \citet[][]{va17info}, which is explained because in that case only a single massive galaxy cluster was studied, and the peak resolution of the simulation was higher ($\approx 30$ kpc), which allowed for more seldom and violent AGN bursts.

In Fig.\ref{fig:evol_cfr_cool} I give the evolution of complexity as a function of cosmic environment for runs Cosmo0cool, Cosmo0AGN and Cosmo0AGN2, compared with the baseline model \footnote{Strictly speaking, an accurate comparison between the baseline non-radiative run and radiative runs is made difficult by a few unavoidable differences in the numerical setups at the start of the simulation: radiative runs adopt a simple run-time prescription for the the re-heating floor by UV radiation associated with reionizaiton \citep[e.g.][]{hm96}, which increases by $\sim 10^2-10^4$ the energy in voids, by raising the temperature of the most rarefied gas from $\sim 1-10 ~\rm K$ (as in non-radiative runs) to 
to $\sim 10^3-10^4 ~\rm K$, which in turn also changes the distribution and strength of accretion shocks \citep[e.g.][]{va09shocks}.  Moreover, in order to attain a realistic $\sim 0.1-1 \rm ~\mu G$  magnetic field level in galaxy clusters after including the additional magnetisation by AGN, in radiative runs the amplitude of the primordial seed magnetic fields is different is  lower than in the baseline run \citep[][]{hack16}. 

The impact of feedback is a sharp increase of complexity 
 $\sim 2-3$ Gyr since the start of the simulation ($z \sim 2-3$)  in both AGN models,  which corresponds to an epoch of outlfows driven by AGN feedback in the simulation, qualitatively consistent with previous work \citep{va17info}.  In the case of the higher power Cosmo0AGN2 model, the complexity within halos is increased compared to the baseline run at most epochs, but  the impact of AGN on cosmic complexity is overall modest 
($\Delta C_{\rm \mu} \leq 30 \%$) if the low and intermediate density environments. This suggests that  the primary driver of complexity in the cosmic web is the shaping of clustering by gravitational interactions (and their associated magneto-hydrodynamical perturbations), while non-gravitational phenomena associated to galaxy formation can only alter this picture at small-scales.}

\subsection{Algorithmic variations to  measure complexity} 

Finally, I tested a few variations in the baseline algorithm to measure complexity, by considering a different time spacing between the simulation output to be considered in order to build the transition matrix probability, $P_{\rm xyz}$, or by varying the number of energy bins adopted to coarse-grain the distribution of the energy fields.

\begin{figure}
\includegraphics[width=0.45\textwidth]{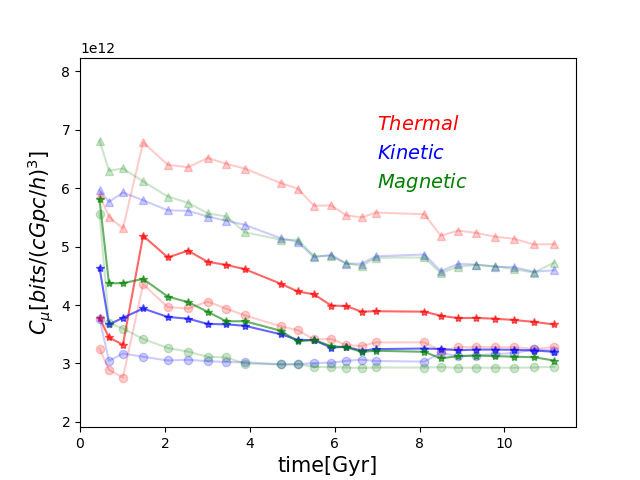}
\includegraphics[width=0.45\textwidth]{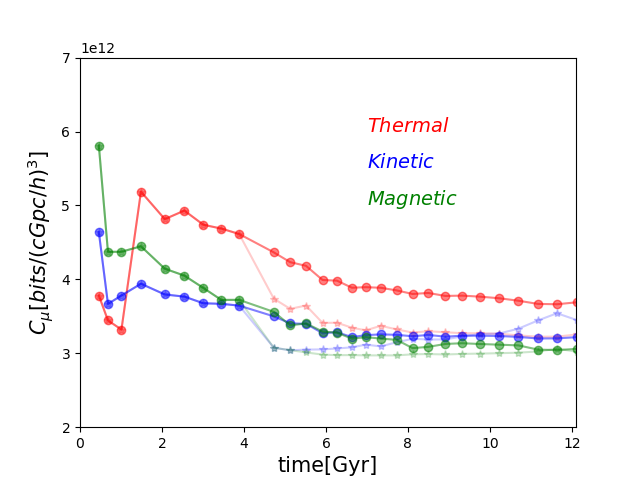}
\caption{Tests on the effect of numerical variations in the algorithm to measure statistical complexity. Top: evolution of $C_{\rm \mu}$ for all energy fields, considering $N_{\rm bin}=50$ (circles), $N_{\rm bin}=100$ (stars) or  $N_{\rm bin}=400$ (triangles) energy bins. 
Bottom: evolution of $C_{\rm \mu}$ using a $dt=5$ separation between timesteps (bold symbols) or $dt=10$ (thin symbols), using an $N_{\rm bin}=200$ binning of the energy fields.}
\label{fig:test}
\end{figure}

The  top panel in Fig.\ref{fig:test} shows the evolution of the total complexity in the volume, by analysing the baseline run with a different number of logarithmic energy bins in Eq.\ref{eq:complex}:  from $N_{\rm bin}=200$ to $N_{\rm bin}=100$ $N_{\rm bin}=50$. 
The bottom panel in the same Figure shows instead the result by  using a different time sampling of timesteps, i.e.  $dt=5$ vs $dt=10$. 

Variations of order $\sim 2$ in the total complexity are measured if the number of energy bins gets reduced by a factor $4$, while the reduction of the time difference between snapshots has a smaller effects on the final distribution of complexity. In both tested variations, the overall trend an relative difference between the complexity of different fields remain fairly constant.
In conclusions, while assessing the convergence of the complexity measurement is made difficult by the fact that the computing effort to measure Eq.\ref{eq:complex} scales as $N_{\rm bin}^2$ (which makes this computation challenging for a large number of energy bins) all trends investigated in the main paper are fairly robust against variations of the proposed algorithm.

\begin{figure*}
\includegraphics[width=0.33\textwidth]{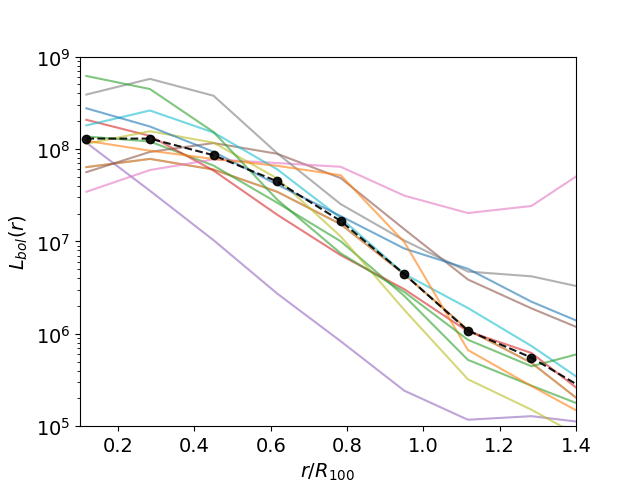}
\includegraphics[width=0.33\textwidth]{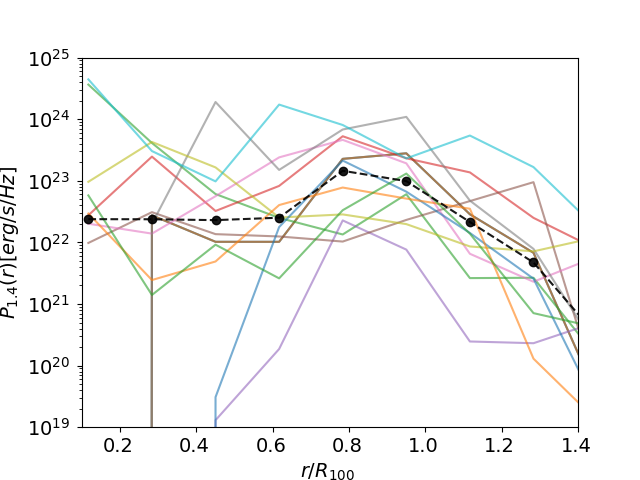}
\includegraphics[width=0.33\textwidth]{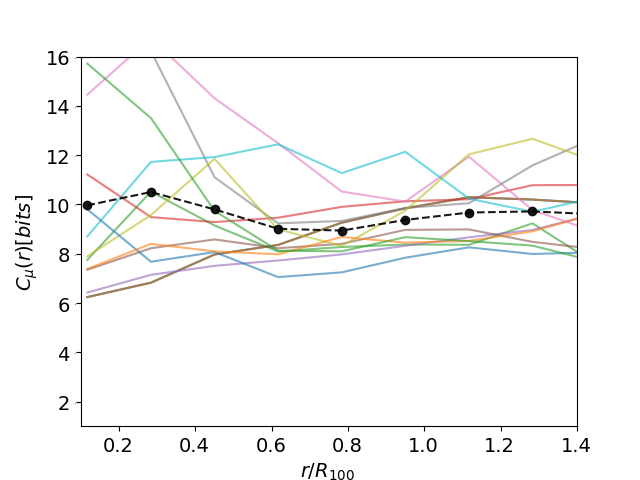}
\caption{Radial profiles of X-ray bolometric emission (in arbitrary units, left), synchrotron radio emission at $1.4$ GHz (centre) and statistical complexity (right) for $M_{\rm 100}\geq 10^{13} M_{\odot} $ halos in run Cosmo0 at $z \approx 0.1$  The colored lines show the average profile of each object (normalized to $R_{\rm 100}$), while the black circles connected by dashed lines give the median profile across the sample.}
\label{fig:prof_clusters}
\end{figure*}

\begin{figure}
\includegraphics[width=0.45\textwidth]{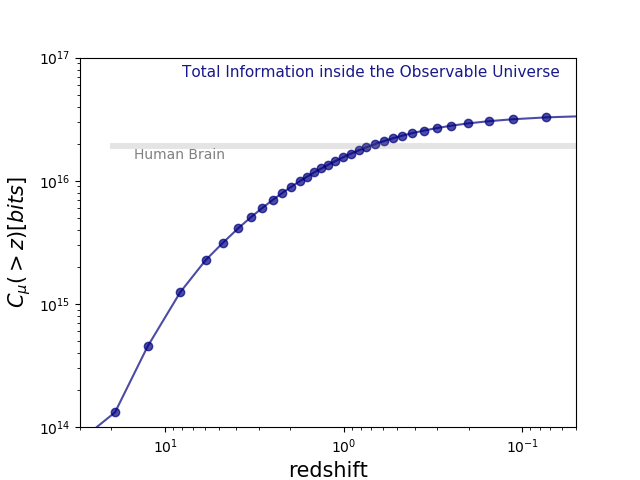}
\caption{Statistical complexity within the observable Universe as a function of redshift (extrapolated from the Cosmo4 run with cosmological parameters from \citealt{2016A&A...594A..13P}). The additional grey line gives the estimated memory capacity of the adult human brain, as a reference.}
\label{fig:universe}
\end{figure}

\section{Conclusions}
\label{sec:conclusions}

Using cosmological numerical simulations  I have investigated the growth of large-scale structures in the cosmic web, and I have studied how physical complexity has emerged starting from simple initial conditions, and under the action of simple physical mechanisms (gravity and magneto-hydrodynamics). 

I have designed an algorithm to measure statistical complexity \citep[e.g.][]{1998PhLA..238..244F,adami,prokopenko2009information} to quantify how much difficult is to predict, in a statistical sense, the evolution of the thermal, kinetic and magnetic energy of baryons in the cosmic web (Sec.\ref{info}), following from my previous work on this subject \citep[][]{va17info}.
The concept of complexity explored in this work is a dynamical one, rather than a geometrical/topological one (the latter approach has been instead applied to studies of the cosmic web using Minkowski functionals and Betti numbers as a proxy for topological persistence of structures, \citealt[e.g.][and references therein]{2017MNRAS.465.4281P}). 

 With the above methodology, derived from Information Theory (already applied first in \citealt{va17info}) it is possible to visualize and measure for the first time (and under certain procedural restrictions) the ubiquitous but very elusive concept of "complexity" and "emergence" in our physical model of the Universe.
 This study  has shown that the environment where the most complex behaviours emerge is the intermediate density regime, associated to  filaments or matter sheets in the
  cosmic web ($1 \leq n/\langle n \rangle \leq 50$) as this environment is interested, at most epochs, by the dissipation of large-scale infall kinetic energy into heating and magnetic field amplification (Sec.\ref{z0}). 

Complexity in the cosmic web has emerged early in time, i.e. already $\sim 2-3 \rm ~Gyr$ after the begin of the simulation, when halos in the cosmic collapsed and converted a large fraction of their gravitational energy into thermal energy and magnetic energy (Sec.\ref{evolution}). This  process was mostly mediated by the formation of violent fluid perturbations (e.g. strong shocks and turbulent motions) which drove large variations in the energy levels of the surrounding intergalactic medium. The amplitude of such variations ubiquitously makes the evolution of gas at intermediate densities  more difficult to predict than in the other more extreme environments. 
  While on small scales the activity of galaxy formation processes (e.g. radiative gas cooling and feedback from active galactic nuclei) can locally introduce more complexity in the evolution of baryons, purely gravitational processes are ultimately setting the overall level of complexity in the entire cosmic web, on its largest scales. 
  
  The variations of the assumed cosmological model, on the implemented gas physics, or the amplitude of cosmic variance are found  only to account for a small, $\leq 20-30\%$ uncertainty 
  in the total cosmic complexity of all energy fields (Sec.\ref{variations}).
  
  All these conclusions have been derived for the fiducial algorithm for statistical complexity presented in Sec.\ref{info}, for the reference spatial resolution of $100 ~\rm kpc/h \approx 140 \rm kpc$, which  in \citet{va17info} was shown to be the best to maximize emergent patterns in the cosmic web. \\
  
The view of complexity of this work radically differs compared to another one often met in the literature, i.e.   what is the {\it maximum} information that the Universe can compute?  The latter is a  question often arising in the holographic description of our Universe \citep[e.g.][for a recent review]{Glattfelder2019}, according to which the total number of degrees of freedom within a finite space-time volume must be proportional to the surface area of the volume \citep[e.g.][]{2004ConPh..45...31B,2005bhis.book.....S}.
Although the theory was originally developed to describe black holes, when applied to the scale of the observable Universe it yields the astounding maximum information capacity of $\sim 10^{100} \rm bits$   \citep[e.g.][]{2003SciAm.289b..58B}, i.e.  orders and orders of magnitude larger than what derived above for the cosmic web. It shall be noticed that while holographic cosmology is concerned with the physical bounds of information that can be contained within the observable Universe, the investigation presented in this paper attempts to constrain the minimum information that characterizes the evolution of the Universe at a specific spatial scales, at which its emergent nature becomes more evident. On these scales, only a limited number of physical mechanisms can add complexity to cosmic evolution (e.g. gravity, fluid-dynamics and cosmic expansion), and thus the information necessary to describe the entire Universe on such large scale is far smaller than the total information that can be enclosed within the same cosmic volume.\\

\subsection{Practical Applications}
It makes sense to ask whether the "cosmic complexity" is a concept of any practical use in astrophysics.
The answer is tentatively positive as this approach offers a quantitative way of constraining which scales and processes must be included in any digital model of the Universe on its largest scales. Assessing what is the minimal model which can reproduce the complexity of the cosmic web of baryons up to the edges of the observable Universe is very functional to
 existing and future large multi-band surveys of the sky wavelengths (e.g. from Euclid to the Square Kilometer Array). The flurry of complex data produced by such surveys call for theoretical models with an equal complexity; the efficient production of such models on a full cosmic scales requires however to careful assess how complex they should really be. \\

From the numerical design viewpoint, the approach of Information Theory explored in this paper may allow numerical codes to identify, at run time, exactly where in the simulation a complex (and thus important) evolutionary pattern is being formed, and thus to selectively concentrate more computing power to better resolve it. Traditionally, this is done using adaptive mesh refinement schemes, in which however the physical conditions to match in order to generate a finer mesh must be set a priori (e.g. refining on matter overdensity, \citealt[e.g.][]{1998ApJ...495...80B,2010MNRAS.401..791S}, on local turbulence conditions, \citealt[e.g.][]{in08}, on shocks, \citealt[e.g.][]{va09shocks}, or local magnetic field values, \citealt[e.g.][]{xu09}, etc). 
Since the metrics of statistical complexity only relies on the symbolic analysis of the numerical datastream, and not on the underlying physics, it can be straightforwardly applied to any  astrophysical simulation (regardless of the specific scale or set of physical processes) in order to refine at run time the level of spatial/time/mass resolution where complex pattern are seen to emerge out of simple initial conditions. 

 Relating complexity with the observable properties of the cosmic web  is probably impossible because statistical complexity is a dynamical measurement, derived for timescales which largely exceed what is accessible to extragalactic observations. However, the association of complexity with frequent matter accretion phenomena in cluster outskirts echoes the fact that an important class of diffuse non-thermal sources in galaxy clusters are indeed preferentially found in their periphery, i.e. "radio relics" sources, which are elongated and steep-spectrum emission regions believed to be associated with cluster merger shocks \citep[e.g.][for a review]{2019SSRv..215...16V}, and potentially the tip of the iceberg of the largest distribution of the "radio cosmic web" \citep[e.g.][]{brown11}. 

To further investigate this, I have computed in Fig.\ref{fig:prof_clusters} the radial profile of bolometric X-ray emission , synchrotron radio emission and statistical complexity for the most massive halos in the baseline run at $z=0.1$. 
While the bolometric X-ray emission is computed as  $L_X \propto n^2 \sqrt{T}$, for the synchrotron radio emission from shock-accelerated relativistic electrons I have use the formalism by \citet{hb07} coupled to a shock-finder for {\enzo} simulations, as detailed in \citet{va15radio}. The radial trend in the profiles well show that the radio emission and the statistical complexity in the simulated clusters have a similarly flat distribution extending out to $\sim R_{\rm 100}$,
where out-of equilibrium fluid perturbations are continuously driven by accreted matter,  while the X-ray emission obviously peaks in the hot and dense core regions, while it gets $\sim 10^2-10^3$ times dimmer in the outer regions of clusters. While resolving in detail the dynamics of intracluster gas at $R_{\rm 100}$ will remain a challange for X-ray observations \citep[e.g.][for recent reviews]{2013SSRv..177..195R,2019arXiv190304550W}, the complex periphery of clusters has become an almost routinely detectable target of radio observations \citep[e.g.][]{2002NewA....7..249B,bo13,2018MNRAS.478..885B,2019Sci...364..981G}. Therefore, existing and future radio observations have the potential of studying a cosmic environment where the most complex phenomena associated with the growth of cosmic structure (still) take place. 
The peripheral radio emission from relics is expected to be just the tip of the iceberg of the larger radio cosmic web, illuminated by structure formation shocks and in the range of what the Square Kilometer Array may detect \citep[e.g.][]{2004ApJ...617..281K,va15ska,va15radio}. On the other hand,  imaging the hot diffuse gas in the cosmic web seems to remain a challenge also in the future  \citep[e.g.][]{2015A&A...583A.142N,2015Natur.528..105E,2019arXiv190910518C,2019arXiv190801778S}.
In summary, while on one hand complexity is likely an unobservable property of cosmic structures, it is useful to theoretically characterize the dynamics of regions which existing and future observation can efficiently target and detect in the radio domain.

  \subsection{How complex is the observable Universe?}

In conclusion, the combination of Information Theory and modern cosmological simulations makes it possible to approach a challenging question: {\it how complex is the Universe we live in}?

Based on the evolutionary trends of complexity measured in  Sec.\ref{evolution},  the statistical complexity within the scale of the observable Universe can be integrated , as:

\begin{equation}
R_c(z)=\frac{c}{H_0} \int_z^\infty \frac{dz'}{E(z')}    
\end{equation}

where $H_0$ is the present-day Hubble constant and $E(z')$ is the cosmological evolution factor, which after the epoch of radiation dominance ($z \leq 3500$) is well approximated as $E(z) \approx [\Omega_M(1+z)^3+\Omega_\Lambda]^{1/2}$ \citep[e.g.][]{2018PASP..130g3001C}.

The average complexity measured at each timestep in the simulated volume (including all three energy fields) can thus be integrated in redshift, which yields the extrapolated total amount of complexity within the radius of the observable Universe today:

\begin{eqnarray}
               C_{\rm Universe}=4 \pi \int_0^{R_c} \langle C_{\rm \mu}(z)\rangle  r^2 dr \label{eq:universe}\\
               = 4 \pi \int_0^{\infty} \langle C_{\rm \mu}(z)\rangle  r^2 \frac{dr}{dz}dz
             \approx 3.56 \cdot 10^{16} \rm bits .
\end{eqnarray}

where the integration is done using the redshift evolution of $C_{\rm \mu}$ measured in the Cosmo4 simulation, based on
the cosmological set of parameters  from  \citet[][]{2016A&A...594A..13P}. The distribution of total complexity as a function of the redshift from $z=30$ to $z \approx 0$ and based on Eq.\ref{eq:universe} is shown in Fig. \ref{fig:universe}.

The total statistical complexity within the  observable Universe at $z \approx 0$ is thus $C_{\rm Universe}=3.56 \cdot 10^{16}$ bits, i.e. to $\approx 4.3$ Petabytes of memory.

 Obviously, this estimate is just a rough indication of the order of magnitude of the amount of information that a numerical model needs to store and process, in order to
 describe the evolution of the gas in cosmic web within the observable Universe on a spatial resolution of $\approx 100$ kpc/h and over a time scale of $\approx 200$ Myr (and for the logarithmic discretization of the gas energy levels as in the algorithm adopted here, e.g. Sec.2.2). Within the range of uncertainties studied in this work, which are connected to cosmic variance, physical prescriptions for  gas physics and uncertainties on the cosmological model, the total "cosmic complexity" should be in the range of $10^{16}-10^{17}$ bits ($\sim 1-10$ Pb) of information.\\
 
To put this number into perspective, it is of the same order of the total amount of data generated every day by social media {\footnote{At least at the time in which this work was written, e.g.  http://res.cloudinary.com/yumyoshojin/image/upload/v1/pdf/future-data-2019.pdf}}.

More interestingly, this number is similar to the latest estimates of the maximum memory capacity of the human brain  (see the additional grey line in Fig. \ref{fig:universe}), which stems from 
the extrapolation of the information that can be stored by synaptic plasticity, translating into a storage capacity of roughly 4.7 bits of information per synapse
\citep[][]{ISI:000373445100001}.
By extrapolating for the total number of neurons and synapses in the human brain, it is estimated that the network connectivity of a typical adult human brain typically stores $\sim 2 \cdot 10^{16} \rm ~bits$, i.e. $\sim 2.5$ Petabytes of data {\footnote{See also http://nautil.us/issue/74/networks/the-strange-similarity-of-neuron-and-galaxy-networks-rp for a recent semi-quantitative comparison between the structural properties of the cosmic web and of the human neuronal network.}}.\\

This similarity further suggests that common analysis techniques and methodologies can be used to study complex networks, despite the entirely different physical mechanisms ruling their evolution.

\section*{Acknowledgements}

I thank the anonymous referee for useful suggestions that have improved the presentation and content of this work.  The cosmological simulations described in this work were performed using the {\enzo} code (http://enzo-project.org), which is the product of a collaborative effort of scientists at many universities and national laboratories. I gratefully acknowledge the {\enzo} development group for providing extremely helpful and well-maintained on-line documentation and tutorials.\\
Most of the analysis done in this work was performed with the Julia code (https://julialang.org). Visualisations in 3D are done using SAO Image ds9 (http://ds9.si.edu/site/Home.html).\\
I acknowledge financial support from the ERC Starting Grant "MAGCOW", no.714196. I acknowledge the  usage of computational resources on the J\"{u}lich Supercomputing Centre (under project HHH42 and STRESSICM), and the  usage of online storage tools kindly provided by the INAF Astronomical Archive (IA2) initiave (http://www.ia2.inaf.it).\\
I thank A. Feletti for useful discussion on the quantitative comparison between the cosmic web and the human brain, and  L. Moscardini and S.Banfi for fruitful scientific discussions on the CDM model and on numerical routines.\\
I finally wish to thank Diletta Vazza, born across the revision of this paper,  for having brought more complexity into my life.

\bibliographystyle{mnras}
\bibliography{info,franco}

\end{document}